\documentclass[flEqn,usenatbib,useAMS]{mnras}

\usepackage{graphicx}	
\usepackage{amsmath}	
\usepackage{amssymb}	
\usepackage{multicol}        

\usepackage[T1]{fontenc}
\usepackage{ae,aecompl}

\usepackage{mathptmx}
\usepackage{txfonts}

\title[Halo dependence on the initial power spectrum]{Connecting the structure of dark matter haloes to the primordial power spectrum}

\author[S. T. Brown et al.]{Shaun T. Brown$^{1}$\thanks{Contact e-mail: \href{S.T.Brown@2018.ljmu.ac.uk}{S.T.Brown@2018.ljmu.ac.uk}}, Ian G. McCarthy$^{1}$\thanks{Contact e-mail: \href{I.G.McCarthy@ljmu.ac.uk}{I.G.McCarthy@ljmu.ac.uk}}, Benedikt Diemer$^{2}$, Andreea S. Font$^{1}$, \newauthor Sam G. Stafford$^{1}$, Simon Pfeifer$^{1}$
\\
$^{1}$Astrophysics Research Institute, Liverpool John Moores University, 146 Brownlow Hill, Liverpool L3 5RF \\
$^{2}$NHFP Einstein Fellow, Department of Astronomy, University of Maryland, College Park, MD 20742, USA
}

\date{date}

\pubyear{year}

\begin{document}
\label{firstpage}
\pagerange{\pageref{firstpage}--\pageref{lastpage}}
\maketitle

\begin{abstract}
A large body of work based on collisionless cosmological N-body simulations going back over two decades has advanced the idea that collapsed dark matter haloes have simple and approximately universal forms for their mass density and pseudo-phase space density (PPSD) distributions. However, a general consensus on the physical origin of these results has not yet been reached. In the present study, we explore to what extent the apparent universality of these forms holds when we vary the initial conditions (i.e., the primordial power spectrum of density fluctuations) away from the standard CMB-normalised case, but still within the context of $\Lambda$CDM with a fixed expansion history. Using simulations that vary the initial amplitude and shape, we show that the structure of dark matter haloes retains a clear memory of the initial conditions. Specifically, increasing (lowering) the amplitude of fluctuations increases (decreases) the concentration of haloes and, if pushed far enough, the density profiles deviate strongly from the NFW form that is a good approximation for the CMB-normalised case. Although, an Einasto form works well. Rather than being universal, the slope of the PPSD (or pseudo-entropy) profile steepens (flattens) with increasing (decreasing) power spectrum amplitude and can exhibit a strong halo mass dependence. Our results therefore indicate that the previously identified universality of the structure of dark matter haloes is mostly a consequence of adopting a narrow range of (CMB-normalised) initial conditions for the simulations. Our new suite provides a useful test-bench against which physical models for the origin of halo structure can be validated.
\end{abstract}


\section{Introduction}
The study of how structure forms in a $\Lambda$-Cold Dark Matter ($\Lambda$CDM) universe has been an active area of research since its general acceptance as the leading cosmological paradigm \citep{Cobe,Riess_1998,WMAP,Planck}. Of particular interest within the large-scale structure of our Universe are collapsed, gravitationally-bound, virialised objects that are commonly referred to as dark matter (DM) `haloes'.

Due to their non-linear evolution, DM haloes are most widely studied in the context of cosmological N-body simulations where the equations of motion are solved explicitly. From numerous such numerical studies there has emerged a few key results that appear to exhibit a degree of {\it universality}.  First, the density profiles of relaxed haloes are well described by a Navarro-Frenk-White (NFW) profile \citep{NFW_96,NFW_97}, defined as 
\begin{equation}
    \rho (r) = \frac{\rho_0}{r/r_s (1+r/r_s)^2}.
    \label{NFW}
\end{equation}
\noindent where $\rho_0$ is a simple normalisation term,\footnote{In fact, the normalisation does not need to be a free parameter, but can be specified by requiring that the integral of the profile matches the total mass of the halo (e.g., within some spherical overdensity radius).} while $r_s$ is a scale radius. The NFW profile has asymptotic inner and outer slopes of $\rho \propto r^{-1}$ and $r^{-3}$, respectively, with $r_s$ being the radius at which the logarithmic slope is $-2$. The scale radius, $r_s$, is often quoted as a concentration parameter, $c \equiv R_\Delta/r_s$, where $R_\Delta$ is the radius containing a mean density of $\Delta$ times the critical (or mean) density of the universe. It has been shown that DM haloes are approximately self similar, as the NFW profile would suggest, however in general DM densities are more accurately described with an additional `shape' parameter \citep{navarro2010}. For this an Einasto profile is often used (see Eqn.~\ref{einasto}) \citep{einasto}.

The second result is that the radial dependence of the pseudo-phase space density (PPSD) profile, $Q(r)$, appears to be well described by a simple power law over radii sampled by simulations (e.g., \citealt{taylor-navarro}). The PPSD is defined as 
\begin{equation}
    Q(r) \equiv \frac{\rho(r)}{\sigma(r)^3},
    \label{ppsd}
\end{equation}
\noindent with $\sigma(r)$ being the total velocity dispersion at a given radii. The exponent of the power law is seemingly a constant for all haloes, with $Q(r) \propto r^{-1.875}$. This suggests that the PPSD profiles are identical for all haloes, potentially indicating that the PPSD is a more fundamental property of DM haloes than the mass density.

Both of these results are well established within the literature, however their physical origins are relatively poorly understood. It has been argued that an inner and outer slopes of the density profile $-1$ and $-3$, respectively, are expected from hierarchical structure formation. Halo assembly can be split into two regimes, an early rapid accretion phase followed by slower accretion. The inner slope of $-1$ is formed during this initial phase and the outer slope of $-3$ during the second phase of slow accretion, with the scale radius being linked to the time a halo transitions between these stages \citep[e.g.][]{Syer&White98,Lu+06}. The power law nature of the PPSD, as well as the specific exponent of $Q(r) \propto r^{-1.875}$, is predicted by secondary infall models \citep{Bertschinger}. However, these models are built on the strong assumption that the initial perturbation and subsequent accretion is spherically symmetric and smooth, i.e. there is no hierarchical growth of structure. Why such a model should correctly match the PPSD profiles of haloes in a `real' universe that grow through the continual anisotropic accretion of clumps of matter is still an open question within the field. Moreover, there is significant scatter on a halo to halo basis as well as some debate over the exact exponent \citep[e.g.][]{ppsd_1.94, faltenbacher_entropy}. It has also been claimed that the PPSD profiles are slowly rolling power laws that only appear to be a perfect power law over the radial ranges sampled by N-body simulations \citep[e.g.][]{ppsd_not_pow}, which is typically above $~1\%$ of a halo's virial radius.

The above discussion applies specifically to a collisionless simulation with no hydrodynamics or non-gravitational physics (e.g., cooling, feedback processes) that are employed in many of the modern galaxy formation and cosmological simulations \citep[e.g.,][]{Illustris,Cosmo_owls,Eagle,apostle,bahamas,FIRE,Illustius_tng}. Additionally, the structure of haloes can be significantly affected by extensions to the $\Lambda$CDM model. This can be done through direct changes to the equations of motion, such as self-interacting dark matter \citep[e.g.][]{Vogelsberger14,Robertson18}, or changes to the linear power spectra in the early Universe, such as warm dark matter \citep[e.g.][]{Apostle_wdm,coco_warm} and a running of the spectral index \citep[e.g.][]{Garrison_2014,Stafford_2020} to name but a few possible extensions (see \citealt{stafford2020exploring} for a recent comparison of the effects of these extensions along with baryonic processes). There are many open questions associated with these more complex simulations, and as such are the focus of much research. However, there is still strong motivation to further study the structure in comparably simple collisionless simulations. If we cannot understand the formation and evolution of haloes in these simulations, it is difficult to see how a rigorous physical picture can be established when the problem is made considerably more complex with additional processes.



The evolution of structure in a cold dark matter universe is driven by three things: i) the initial conditions, i.e. the primordial power spectra; ii) the nature of the gravitational force law; and iii) the cosmological parameters that, with the gravitational force law, determine the expansion history of the universe. In the present study, we leave the force law and expansion history unchanged from the accepted $\Lambda$CDM model and focus solely on the effects of varying the primordial power spectrum on the properties of late-time collapsed haloes. This is achieved by systematically varying the primordial spectral index, $n_s$, and amplitude, $A_s$, to study both the effects of shape and amplitude changes in the initial power spectra.  We note that it has previously been shown that the NFW form and power law PPSD generally continue to be accurate descriptions of the structure of DM haloes for a wide range of changes to the shape of the linear power spectra and to the cosmological parameters that control the expansion history (e.g., \citealt{NFW_97,NFW_indepent_ICs1,NFW_indepent_ICs2,NFW_indepent_ICs3}) and even for some setups that avoid hierarchical clustering altogether (e.g., \citealt{Huss1999}).   It is interesting to note, however, that previous studies did not explore a wide range of primordial power spectrum amplitudes, which we find to be key.

Using cosmological simulations, we show that many of the results discussed above are only valid in universes close to our own, specifically for universes with a similar amplitude of initial density fluctuations. Varying away from the standard initial conditions, we find that the slope of the PPSD is in general not a constant and instead exhibits a clear mass dependence. Furthermore, the NFW form becomes an increasingly poor description of the mass structure of haloes as the amplitude of the primordial power spectrum is increased.

The paper is organised as follows. In Section~\ref{technical_section} we discuss the technical details of the simulations and provide motivation for the choice of parameters. In Section~\ref{results} we present stacked density, entropy and velocity profiles of haloes in the various simulations. In Section~\ref{mass_dependence_peak_height} we study how fitted density and entropy parameters vary with mass and peak height. In Section~\ref{MAH_CMH} we present how the accretion histories of haloes vary in the cosmologies studies here as well as comparing our results for the concentration--mass (or similarly concentration--peak height) relation to the predictions of a number of (semi-)analytic models. Finally, in Section~\ref{summary} we summarise our findings.

\section{Simulations and halo property estimates} \label{technical_section}

In this section we describe the general simulation setup and the generation of initial conditions. We also describe our halo selection criteria and how halo properties are estimated.

\subsection{Halo mass and radius definitions}

Throughout this paper we define the size and mass of haloes as a spherical overdensity, $\Delta$, with respect to the critical density of the universe. The labels $M_{\rm{200c}}$ and $R_{\rm{200c}}$, as we use in this work, therefore correspond to the mass and size, respectively, of haloes defined such that the average density within $R_{\rm{200c}}$ is equal to $200$ times the critical density.

\subsection{General simulation setup}

All simulations used in this work share the same technical details. Specifically, the linear power spectra that are used to generate the initial conditions (see Section \ref{ICs_section}) are computed using the Boltzmann code \texttt{CAMB} \citep{camb}, at a starting redshift of $z=127$. The initial particle positions and velocities are then generated using a modified version of \texttt{N-GENIC}\footnote{The publicly available version of this code can be found at \url{https://github.com/sbird/S-GenIC}.} \citep{Gadget_2}, including second-order Lagrangian perturbation theory (2LPT) corrections and adopting identical phases for all simulations. The simulations use a comoving $200$ $h^{-1}$ Mpc on a side box with $512^3$ particles. For the background cosmology we use the best fit WMAP 9-yr results \citep{hinshaw2013}, with $h=0.7$, $\Omega_m=0.2793$, $\Omega_b=0.0463$ and $\Omega_{\Lambda}=0.7207$. As such the particle masses of all these simulations is $4.62 \times 10^{9}h^{-1}$ M$_{\odot}$. The simulations are run with a modified version of the \texttt{GADGET-3} code \citep{Gadget_2,bahamas}. The gravitational softening is fixed to $4$ $h^{-1}$kpc (in physical coordinates for $z \leq 3$ and in co-moving at higher redshifts). In Appendix~\ref{Resolution_test} we present a resolution and box size study for a range of our cosmologies to make sure these numerical parameters do not effect the results presented in this paper.

All haloes are identified with the \texttt{SUBFIND} algorithm \citep{Subfind}. In this work we will only present results of host haloes, which are defined as the largest halo in the friend-of-friends (FOF) group. The halo finder here is only used to initially find the FOF group, provide the location of the centre of potential and bulk properties such as $M_{\rm{200c}}$ and $R_{\rm{200c}}$.

\begin{figure*}
    \centering
    \includegraphics[width=2\columnwidth]{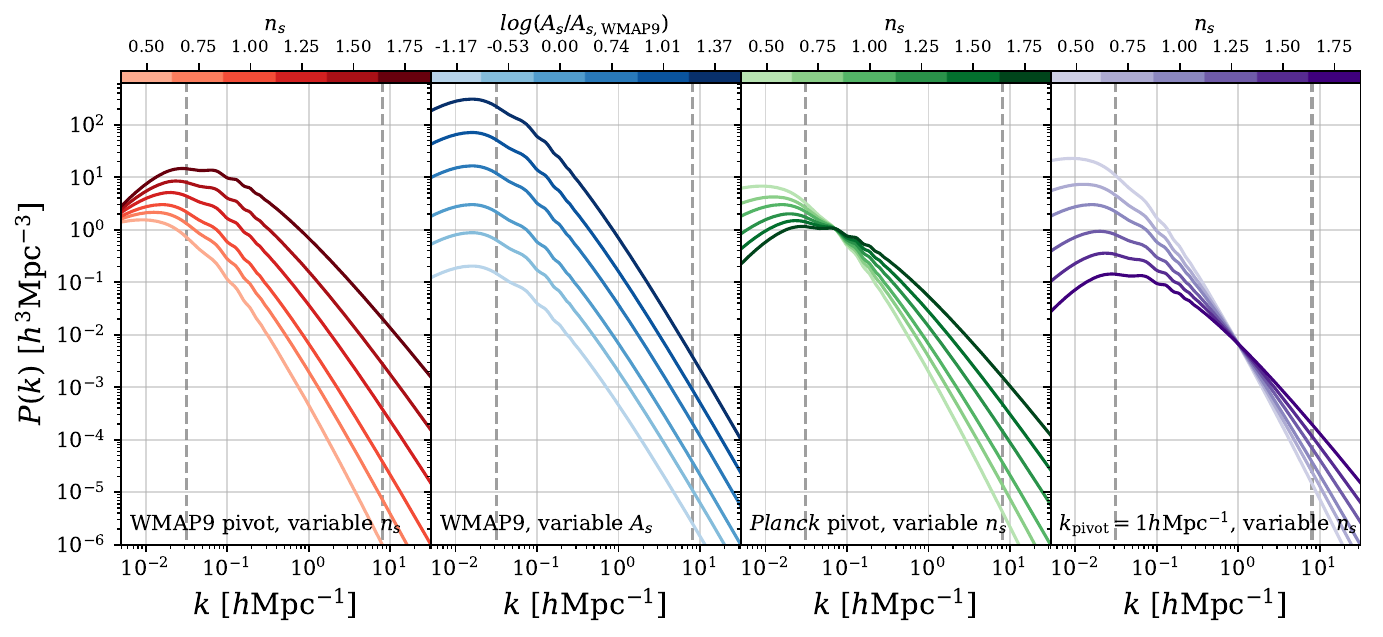}
    \caption{Linear power spectra used to create the initial conditions of the various simulations at $z=127$, generated using CAMB \citep{camb}. All cosmologies have the same background cosmology (best fit WMAP 9-yr results), but with variable $A_s$, $n_s$ and $k_{\rm{pivot}}$. Left panel shows the fiducial suite with adopting a WMAP pivot point ($k=2.85 \times 10^{-3} h \text{Mpc}^{-1}$) with systematically varying $n_s$. The middle-left panel is the initial power spectra for the suite of simulations that have a fixed $n_s$ but a varied $A_s$ such that at $k=1 h \text{Mpc}^{-1}$ they have the same power as the corresponding WMAP pivot simulations (left panel). The middle-right panel shows the same but for a \textit{Planck} pivot point ($k=7.14\times 10^{-2} h \text{Mpc}^{-1}$). The suite using a $k_{\rm{pivot}}=1h$ Mpc$^{-1}$ with systematically varied $n_s$ is shown in the right most panel. The value of $n_s$, or equivalently matched $A_s$ is shown in the colour bar above each plot. The vertical dashed lines represent the Nyquist frequency and the fundamental mode for our box size, i.e. the resolved range of our simulations.}
    \label{fig:initial_pow}
\end{figure*}

\subsection{Primordial power spectra} \label{ICs_section}

We study the effect of the initial density field by varying two free parameters in the $\Lambda$CDM model that directly affect the primordial density perturbations, $A_s$ and $n_s$. Here, $n_s$ is referred to as the primordial scalar spectral index and it is the slope of the primordial power spectrum, assumed to be a power law. $n_s$ is a free parameter in the $\Lambda$CDM model that is well-constrained by the CMB to have a value of $n_s \approx 0.96$--$0.97$ \citep{WMAP, planck2015}. $A_s$ is the amplitude of the initial power spectrum, which is specified at a chosen pivot point, $k_{\rm pivot}$. Systematically varying these two parameters, as well as the pivot point, allows us to isolate shape and amplitude changes to the initial power spectrum, but without changing the background expansion or the nature and abundance of dark matter. Thus, our simulations are still run in the context of $\Lambda$CDM.

To study the dependence of $n_s$ and $A_s$ on the properties of DM haloes we initially take a WMAP 9-yr cosmology and systematically vary $n_s$ between 0.5 to 1.75. While all of the runs we consider (apart from the fiducial $n_s=0.96$ case) are ruled out by observations, our goal here is to see whether and to what extent the properties of collapsed haloes `remember' the initial conditions.

The initial linear power spectra at $z=127$ for this suite of simulations are shown in the left panel of Fig.~\ref{fig:initial_pow}. The change in $n_s$ can clearly be seen in the slope of the power spectra at small k-modes ($k \sim 10^{-2}$ $h$ Mpc$^{-1}$), with larger $n_s$ values giving steeper profiles and vice versa. Another notable feature is the pivot point, in this case $k_{\rm{pivot}}=2.86\times10^{-3} h~{\rm Mpc}^{-1}$ (or $2\times10^{-3}$ Mpc$^{-1}$), where the power spectra are normalised. When looking at the region sampled by our simulations (delineated by the vertical dashed lines), it becomes clear that the power spectra will depend on both $n_s$ and $A_s$, as well as the choice of $k_{\rm{pivot}}$. $n_s$ can be interpreted as a shape change but the change in amplitude at a given $k$ scale is a combination of $n_s$, $A_s$ and $k_{\rm{pivot}}$. Due to this complexity, we isolate the amplitude and shape changes by running three complimentary sets of simulations: (i) a set of simulations where $n_s$ is fixed to $0.96$ but the amplitude $A_s$ is varied, such that the power at $k=1 h~{\rm Mpc}^{-1}$ is equivalent to the WMAP 9-yr simulations with varied $n_s$ (see middle-left panel of Fig.~\ref{fig:initial_pow}); (ii) a set of simulations using the same pivot point as the $\textit{Planck}$ team, $k_{\rm{pivot}}=7.14 \times 10^{-2} h~{\rm Mpc}^{-1}$, and again systematically vary $n_s$ (see middle-right panel of Fig.~\ref{fig:initial_pow}); and (iii) a set of simulations where we use $k_{\rm{pivot}}=1 h~{\rm Mpc}^{-1}$ and again systematically vary $n_s$ (see right panel of Fig.~\ref{fig:initial_pow}). In this way we have two suites of simulations with a combination of shape and amplitude changes, but to varying degrees, and additionally two sets of simulations where we have tried to isolate the changes in amplitude and shape. For two suites we have specifically emphasised $k=1 h~{\rm Mpc}^{-1}$, either by using it as the pivot point or specifically matching the amplitude at that k-scale. We have chosen this particular $k$-mode as it is well sampled in our box, well away from the Nyquist frequency and far from cosmic variance at larger $k$-modes, as well as being a $k$ scale associated with haloes of $M_{\rm{200c}} \sim 10^{13-14} h^{-1}M_{\odot}$ \citep{vanDaalen2015}, which is the mass of a typical halo in our simulations.

While different power spectra are often normalised by $\sigma_8$ \citep[e.g.][]{Knollmann2008}, in this work we choose to normalise at a particular $k$-scale, $k_{\rm{pivot}}$. Using a pivot scale allows us to isolate the effects of amplitude and shape changes to the initial power spectrum by varying $k_{\rm{pivot}}$. The use of a pivot point also allows a more intuitive link between a halo of a certain mass and the initial linear power spectrum. Systematically varying the pivot point is similar, though not identical, to choosing to normalise by differing values of $\sigma(R)$ rather than $\sigma(R=8h^{-1}\text{Mpc})$ specifically.

A list of the various cosmologies and simulations parameters is given in Table.~\ref{simulation_table}. The format used in Fig~\ref{fig:initial_pow} to represent the different simulations (such as the colour, panel position, etc.) is the same throughout the paper. Throughout the paper we will use $n_s$ as a reference to a particular simulation regardless of whether $n_s$ is the direct cause of effects we observe. When describing the effect of changing the slope of the power spectrum, which is a change of $n_s$, we will specifically describe it as a slope or general shape change, $n_s$ is simply a reference to the particular simulation being discussed.

\begin{table}
\caption{Summary of the various cosmological parameters for all simulations presented in this work. The main two parameters varied are $n_s$ and $A_s$. Along with $k_{\rm{pivot}}$, they completely specify the initial power spectrum. Note that for the `Matched amplitude' suite, the shape is fixed ($n_s=0.96$) while the amplitude at a scale of $1~h$ Mpc$^{-1}$ is adjusted to match the various runs in the `WMAP9 pivot' suite. All cosmologies have the same background expansion: $h=0.7$, $\Omega_m=0.2793$, $\Omega_b=0.0463$ and $\Omega_{\Lambda}=0.7207$.}
\label{simulation_table}
\begin{tabular}{lllllllll}
\hline
Simulation suite                                 & $n_s$  & $A_s$ [$10^{-9}$]                 & $k_{\rm{pivot}} [h$~Mpc$^{-1}]$ & $\sigma_8$  \\
\hline
WMAP9 pivot                  & $0.96$ & $2.392$   & $2.86 \times 10^{-3}$         & $0.801$               \\
WMAP9 pivot                   & $0.5$  & $2.392$                       & $2.86 \times 10^{-3}$                            & $0.328$                 \\
WMAP9 pivot                  & $0.75$ & $2.392 $                      & $2.86 \times 10^{-3}$                            & $0.530$                \\
WMAP9 pivot                  & $1.25$ & $2.392 $                      & $2.86 \times 10^{-3}$                            & $1.442$                \\
WMAP9 pivot                   & $1.5$  & $2.392 $                       & $2.86 \times 10^{-3}$                            & $2.422$        \\
WMAP9 pivot                  & $1.75$ & $2.392 $                      & $2.86 \times 10^{-3}$                            & $4.114$             \\
Matched amplitude            & $0.96$ & $0.1617 $  & $2.86 \times 10^{-3}$                            & $0.208$              \\
Matched amplitude           & $0.96$     & $0.6992 $ & $2.86 \times 10^{-3}$                           & $0.433$             \\
Matched amplitude           & $0.96$     & $13.08 $  & $2.86 \times 10^{-3}$                           & $1.874$             \\
Matched amplitude            & $0.96$     & $24.45 $  & $2.86 \times 10^{-3}$                            & $3.897$            \\
Matched amplitude           & $0.96$     & $56.55 $  & $2.86 \times 10^{-3}$                           & $8.103$           \\
\textit{Planck} pivot                    & $0.5$  & $2.103 $                     & $7.14 \times 10^{-2}$         & $0.687$             \\
\textit{Planck} pivot                   & $0.75$ & $2.103 $                      & $7.14 \times 10^{-2}$                           & $0.743$           \\
\textit{Planck} pivot                   & $1.25$ & $2.103 $                      & $7.14 \times 10^{-2}$                            & $0.904$            \\
\textit{Planck} pivot                    & $1.5$  & $2.103 $                      & $7.14 \times 10^{-2}$                            & $1.016$             \\
\textit{Planck} pivot                   & $1.75$ & $2.103 $                      & $7.14 \times 10^{-2}$                            & $1.154$      \\   
$k_{\rm{pivot}}=1~h$ Mpc$^{-1}$  & $0.5$  & $1.892$                      & $1.00$                           & $1.261$          \\
$k_{\rm{pivot}}=1~h$ Mpc$^{-1}$ & $0.75$ & $1.892 $                      & $1.00$                            & $0.980$           \\
$k_{\rm{pivot}}=1~h$ Mpc$^{-1}$ & $1.25$ & $1.892 $                      & $1.00$                            & $0.616$            \\
$k_{\rm{pivot}}=1~h$ Mpc$^{-1}$  & $1.5$  & $1.892$                      & $1.00$                            & $0.498$            \\
$k_{\rm{pivot}}=1~h$ Mpc$^{-1}$ & $1.75$ & $1.892 $                      & $1.00$                            & $0.407$          \\

\hline
\end{tabular}
\end{table}

\subsection{Halo selection criteria}

To ensure the robustness of our results, we will focus on well-resolved, relaxed haloes. The first cut made to the halo sample is to remove haloes that are too poorly sampled to generate reliable density and velocity dispersion profiles. This cut is such that the number of particles within $R_{\rm{200c}}$, $N_{\rm{200}}$, is greater than $2 \times 10^3$. This choice results in converged density and velocity dispersion profiles and also avoids some systematic issues observed when fitting poorly sampled haloes (see Appendix \ref{appendix_dens_method} for further details). The second cut is to remove unrelaxed haloes from the sample. For this we use one of the criteria advocated in the work of \citet{neto}. Specifically, that the normalised offset of the centre of mass (CoM) to centre of potential (CoP) is $s=(r_{\rm{CoP}}-r_{\rm{CoM}})/R_{\rm{200c}} <0.07$. \citet{neto} applied additional cuts to the relative mass in substructure and virial ratio, however \citet{duffy} found that a simple cut on CoM and CoP offsets is sufficient to remove the majority of unrelaxed haloes. The fraction of haloes removed from the sample due to this relaxation criteria varies from simulation to simulation but is typically in the range $5$--$10 \%$ for most of the simulations, with the exception of the two simulations with the smallest amplitudes (the lightest red and blue lines in Fig.~\ref{fig:initial_pow} corresponding to the cosmologies with $\sigma_8 = 0.328$ and $0.208$, see Table~\ref{simulation_table}) where around $30 \%$ of haloes are discounted. We have included the relaxation cut to be consistent with other work in the literature, however, we find that there is little to no effect if we include the relaxation cuts or not. We attribute this to us exclusively studying median stacked profiles throughout this work as opposed to individual haloes (see Section~\ref{mass_dependence_peak_height}).

\subsection{Density and velocity dispersion profiles} \label{density and velocity calculation}

We introduce a new method to measure smoothed density and velocity dispersion profiles that have reduced noise and fewer systematic errors compared to the standard method of radial binning. In this section we will outline the general method. An in depth discussion can be found in Appendix \ref{appendix_dens_method}.

The general procedure is to use a weight function to calculate the spherically-averaged density and velocity dispersion of a DM halo. The density is calculated as 
\begin{equation}
    \rho (r) =  \sum^{N_{\rm{kern}}}_i W(r_i;h,r) \enskip m_i .
\end{equation}
We use $r_i$ to denote particle position whereas $r$ is the radius at which the density is being estimated. Here we have written it in a generalised form with variable mass, $m_i$, but for our simulations this is a constant. $W(r_i)$ is a general weight function, which we will define shortly. Bulk velocities, in the three spherical directions, are calculated as 
\begin{equation}
\mathbf{v}_{\rm{bulk}}(r)= \frac{\sum^{N_{\rm{kern}}}_i W(r_i;h,r) \enskip \mathbf{v}_i}{ \sum^{N_{\rm{kern}}}_i W(r_i;h,r)}
\end{equation}
and from this the velocity dispersion is calculated as 
\begin{equation}
\mathbf{\sigma}^2(r)= \frac{\sum^{N_{\rm{kern}}}_i W(r_i;h,r) \enskip (\mathbf{v}_i-\mathbf{v}_{\rm{bulk}})^2}{ \sum^{N_{\rm{kern}}}_i W(r_i;h,r)}.
\end{equation}
In this manner we calculate the velocity dispersion in all three orthogonal directions. Throughout this work we present results for the total velocity dispersion, $\sigma^2=(\sigma^2_{r}+\sigma^2_{\theta}+\sigma^2_{\phi})/3$, and the velocity anisotropy averaged over both angular directions, $\beta=1-0.5\sigma_{\theta}^2/\sigma_{r}^2-0.5\sigma_{\phi}^2/\sigma_r^2$.

The weight function we adopt is the cubic spline implemented in many smoothed particle hydrodynamics methods \citep{Cubic_spline}:
\begin{equation}
    W(r_i;h,r)=\frac{1}{\pi h (0.25 h^2+3r^2)}
    \begin{cases}
    1-6 x^2+6  x ^3 & \text{if } x<0.5\\
    2 (1-x)^3 & \text{if } 0.5<x<1\\
    0 & \text{otherwise }
    \end{cases}
    \label{weight_function}
    ,
\end{equation}
where we have let $x=\frac{|r_i-r|}{h}$. The normalisation factor is only relevant for calculating the density and is the effective volume of the kernel (equivalent to the factor of $\frac{4}{3} \pi [(r+h)^3-(r-h)^3]$ for a square or top hat kernel).\footnote{The normalisation term is different from the usual factor of $8/(\pi h^3)$ as this equation is designed to calculate the mass per unit volume from radial coordinate.}

As can be seen in Eqn.~(\ref{weight_function}) this method has one free parameter $h$, the width (or `smoothing length') of the kernel. We vary $h$ as a function of radius to be equivalent to using logarithmically spaced bins, meaning that $h(r) = A r$. We then let $A$ be a function of $N_{\rm{200c}}$, the number of particles within $R_{\rm{200c}}$. This is done in such a way that the two types of error associated with this method, random Poisson errors from having a finite number of data points and systematic errors from having a kernel with a finite width, scale the same with resolution. The relation used is then
\begin{equation} \label{kernel_width_scaling}
    h(r)=(N_{\rm{200c}}/500)^{-1/3} r.
\end{equation}
where $N_{\rm{200c}}^{-1/3}$ is derived analytically and the factor of $500$ found empirically. In general particles are not required to be uniquely associated with a particular `bin' and kernels can overlap. This allows the kernel width and sampling positions to be independent, which is of particular importance when taking derivatives of the density profiles as we do in this work.

\begin{figure*}
    \centering
    \includegraphics[width=2\columnwidth]{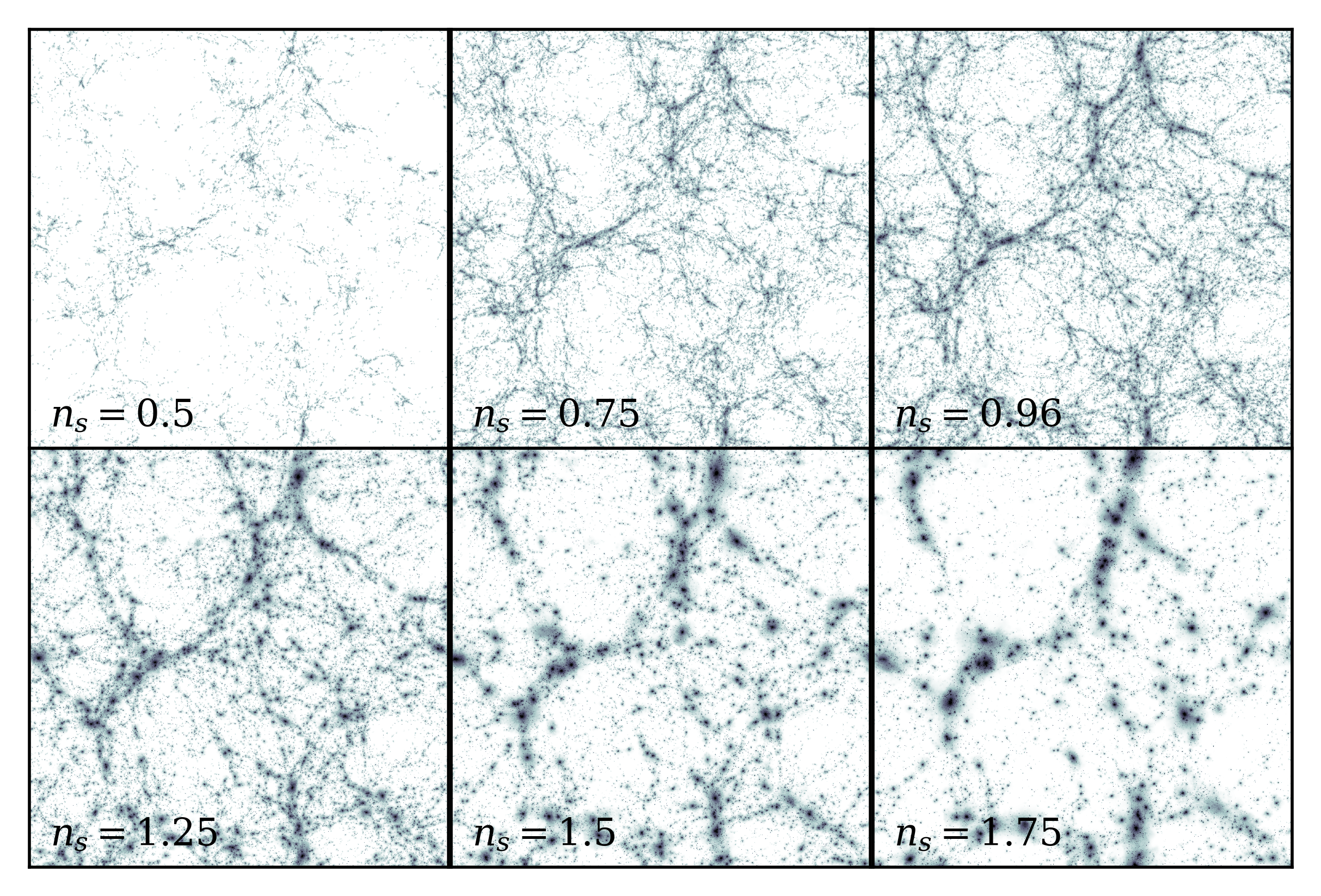}
    \caption{Surface density maps of the six WMAP pivot cosmologies with variable $n_s$ at z=0. From top left to bottom right we have increasing values of $n_s$ from $0.5$--$1.75$, see label in bottom left of each panel. These plots are meant to show the qualitative behaviour of the different density fields so no colour bar is given. Each colour map is normalised the same and represents the logarithmic projected surface density. These images have been made using the publicly available code \texttt{SPH-Viewer} \citep{SPH_viewer}.}
    \label{dens_map}
\end{figure*}

\subsection{Halo entropy vs.~PPSD}

Before proceeding further let us discuss PPSD a little more, particularly why it is referred to as \textit{pseudo phase-space density}. The quantity defined in Eqn.~(\ref{ppsd}) is referred to as PPSD as a simple analogy to true phase-space density as both quantities have the same dimensionality. We believe, however, that there is a better analogous property to be compared to, that of entropy. Taking the definition of entropy in an astrophysical context\footnote{Boltzmann entropy and `astrophysical' entropy are related by the following, $S_{\rm{Boltz}}=ln(S_{\rm{Astro}}^{3/2})+const$.} (e.g., as often employed in studies of the X-ray emission of galaxy clusters), $S \equiv k_B T n_e^{-2/3}$, and making the substitution $T \rightarrow \sigma^2$ and $n_e \rightarrow \rho$, the `entropy' can be written as 
\begin{equation}
\label{Entropy}
    S = \bigg ( \frac{\rho}{\sigma^3} \bigg )^{-2/3}= Q^{-2/3}.
\end{equation}
If the PPSD profiles are power laws with an exponent $-1.875$ then the implied entropy profiles will be power laws with an exponent of $\approx1.25$. The entropy profiles of both real and simulated clusters (this being true entropy of the hot intracluster gas) can be well fit by a power law at large radii, with observed exponents of $\sim 1.1-1.3$ (e.g., \citealt{Voit2005,McCarthy2008,Cavagnolo2009}), suggesting this analogy is not completely unwarranted. Strictly speaking, the quantity in Eqn.~(\ref{Entropy}) should be referred to as \textit{pseudo-entropy}; as we will not be presenting any results from hydrodynamic simulations we simply refer to it as entropy without the risk of confusion. Throughout the rest of this paper we refer solely to entropy and not the PPSD, noting that the former can be trivially converted to the latter as described above.

\begin{figure*}
    \centering
    \includegraphics[width=2\columnwidth]{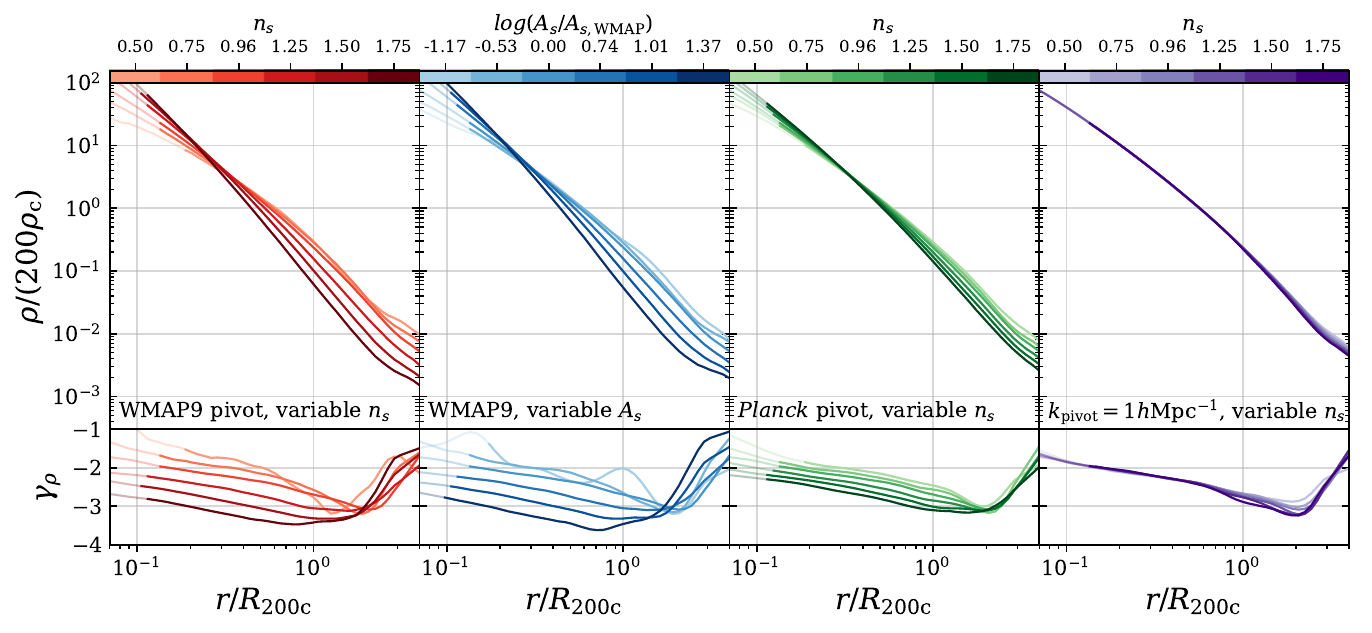}
    \caption{Top panels show the stacked median density profiles of haloes in the mass range $M_{\rm{200c}}=10^{13}$--$10^{13.5} h^{-1}$ M$_{\odot}$ while the bottom panels the logarithmic slope of the above plot, defined as $\gamma_{\rho}=d(\ln\rho)/d(\ln r)$ at $z=0$. Densities are normalised to the critical density today while the radii are normalised to each haloes $R_{\rm{200c}}$. The left column shows the WMAP pivot cosmology with variable $n_s$, the middle-left panels the equivalently matched $A_s$ with fixed $n_s$, the middle-right panels the \textit{Planck} pivot point with variable $n_s$ and the far right panels the $k_{\rm{pivot}}=1h$ Mpc$^{-1}$ suite. The colours represent the different suites of simulations while the shade represents the value of $n_s$ or matched $A_s$, see colour bar above each plot. Each curve is plotted transparent where the criteria for convergence is not met.  A comparison of the different panels indicates that the dominant factor in setting the density profile of a halo is the {\it amplitude} of the linear power spectra at an associated $k$-mode.}
    \label{fig:dens_profile}
\end{figure*}

\section{Stacked profiles} \label{results}

In this section we present the stacked density, entropy and velocity profiles of haloes within a narrow mass range at $z=0$ for the various simulations. An alternative to studying haloes at fixed mass and redshift would be to instead look at haloes as a function of `peak height', as we discuss in Section~\ref{Peak_height_dependence}. The discussion in this section will mainly be limited to describing the recovered features and trends, leaving interpretation of the results until Sections~\ref{mass_dependence_peak_height} \& \ref{MAH_CMH}.

In Fig.~\ref{dens_map} we show the projected density maps for the WMAP 9-yr with variable $n_s$ suite of simulations. We present this plot to highlight in a qualitative way the effect that varying this parameter has on the overall present-day density field.  It is readily apparent that the size and mass of haloes in the various simulations are extremely different, with $n_s=0.5$ (top left) having a largest halo of only $M_{\rm{200c}}\sim 10^{13} h^{-1}$ M$_{\odot}$ while $n_s=1.75$ has haloes that exceed $10^{16} h^{-1}$ M$_{\odot}$. 

\subsection{Stacked density profiles} \label{density_section}

In Fig.~\ref{fig:dens_profile} we present the stacked density profiles of haloes in the mass range $M_{\rm{200c}} = 10^{13}\text{--}10^{13.5} h^{-1} M_{\odot}$ at $z=0$. Haloes are stacked by taking the median densities of individual haloes in units of $r/R_{\rm{200c}}$. The top panels show the density as a function of radius, with density normalised by $200\rho_{\rm{c}}(z=0)$ and radii scaled by $R_{\rm{200c}}$ of each halo (i.e., we stacked in bins of normalised radius). All profiles are plotted out to $4R_{\rm{200c}}$. In the bottom panels we plot the corresponding logarithmic slopes, defined as $\gamma_{\rho} = d(\ln{\rho})/d(\ln{ r})$. The different columns represent the different sets of simulations: left (in red) is the WMAP pivot point with variable $n_s$; middle-left (in blue) is the WMAP pivot with fixed $n_s$ but variable $A_s$ to match the amplitude of the equivalent $n_s$ run at $k=1~h$ Mpc$^{-1}$; middle-right (in green) is the \textit{Planck} pivot with variable $n_s$; and right (in purple) is the case with variable $n_s$ and a pivot scale of $k_{\rm{pivot}}=1~h$ Mpc$^{-1}$. The shade of the colour represents the value of $n_s$ used (or amplitude matched to) with darker colours corresponding to higher values of $n_s$ (or higher matched $A_s$), see colour bars above each column. The colour schemes are equivalent to Fig.~\ref{fig:initial_pow} so can be directly compared. We recommend the reader refer back to Fig.~\ref{fig:initial_pow} for intuition of what part of the initial power spectra has changed. 

Let us first focus our attention on the left column of Fig.~\ref{fig:dens_profile}, which we use as the fiducial suite of simulations. The first thing to notice is the strong dependence that the steepness of the density profiles at a fixed $r/R_{\rm{200c}}$ has on $n_s$, simulations with larger values of $n_s$ result in more negative logarithmic slopes. It appears that the main difference between these density profiles is a change in concentration or similarly scale radius. Looking at the logarithmic slope of these profiles (bottom panel), the profiles all behave roughly log-linearly before a sharp decrease at $\sim R_{\rm{200c}}$. This feature in the density profile corresponds to the splashback radius, the radii at which particles reach the apocenter of their first orbit. The splashback radius roughly delineates a region that is actively accreting onto the halo from the background universe. The behaviour and shape of $\gamma_{\rho}(r)$ in this region is in qualitative agreement with other work \citep[e.g.][]{splashback14,Adhikari2014,More2015}. 

At all radii within the splashback radius, we see that larger values of $n_s$ lead to steeper density slopes. Recall that these plots are at fixed mass and redshift, we are therefore not studying haloes of the same age or equivalent accretion histories. The results are therefore not necessarily directly due to the change in the slope of the initial power spectra, as we discuss shortly.

It appears that the differences in logarithmic slope for different values of $n_s$ are primarily just an amplitude offset, suggesting that the density profiles remain approximately self similar but with different concentrations. Further out, near and beyond the splashback radius, there is a much more complicated dependence on $n_s$. For example, the radius at which the splashback feature occurs (i.e. the minima of $\gamma_{\rho}$) seems to be a non-monotonic function of $n_s$. Similarly, the logarithmic slope beyond the splashback radius has a complicated dependence on $n_s$ with no easily discernible trend, although there are weak hints that the logarithmic slope past the splashback radius grows more slowly for larger $n_s$.

\begin{figure*}
    \centering
    \includegraphics[width=2\columnwidth]{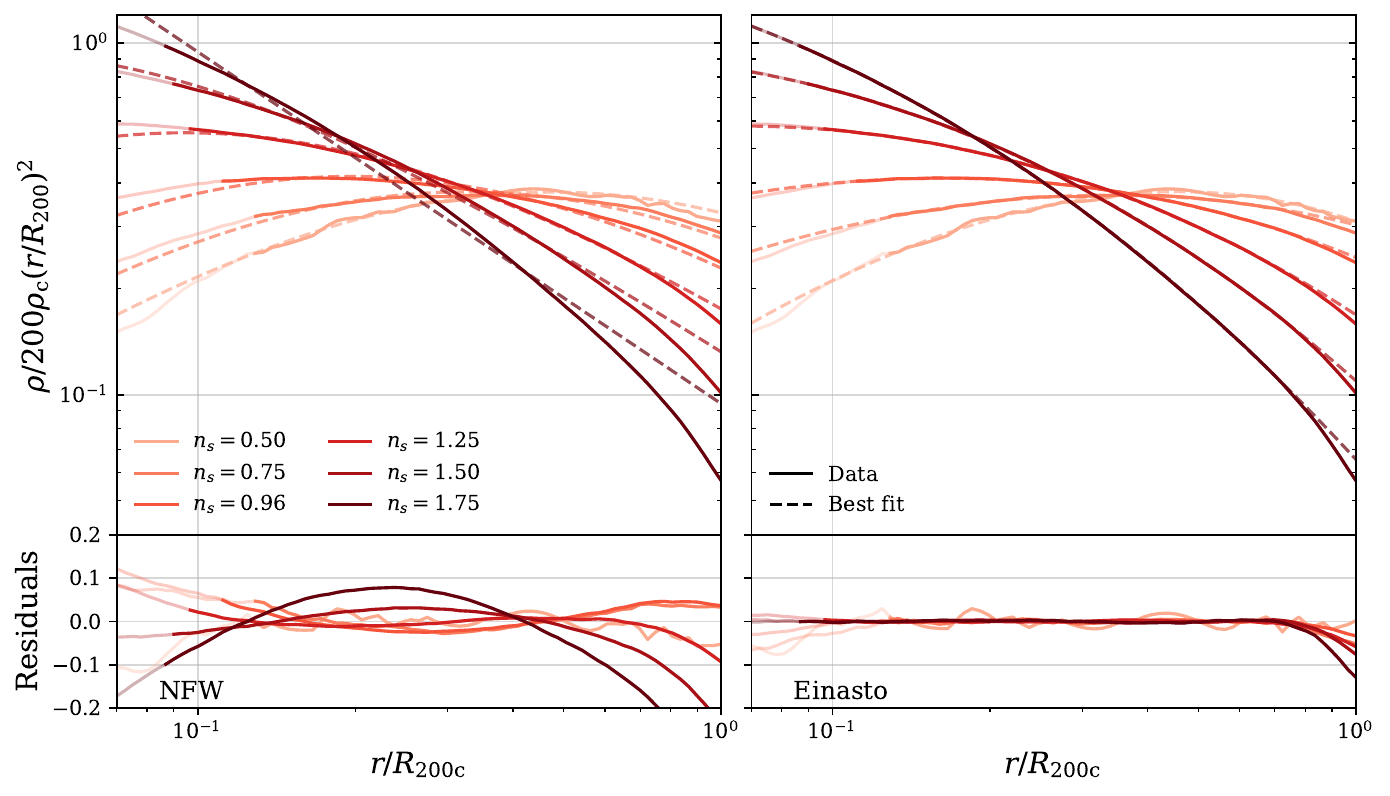}
    \caption{Top panels show the density profiles of the WMAP pivot cosmology with variable $n_s$, see legend. This is identical to the top left panel of Fig.~\ref{fig:dens_profile}, apart from multiplying by $(r/R_{\rm{200c}})^2$ to reduce the dynamic range. We have then fit each profile with either a NFW (left) or Einasto curve (right), allowing both the scale radius and shape parameter to vary when fitting the Einasto profile. The fitting routine minimises our figure of merit, $\psi^2$, and is only fit over the range where $r/R_{\rm{200c}}<0.7$, to avoid fitting to the splashback radius, and above the convergence radius. On the bottom panels we show the fractional residuals, $(\rho_{\rm{fit}}-\rho)/\rho$, from this fit.  The NFW form breaks down for runs with large $n_s$ (corresponding to large amplitudes at $k=1 h$ Mpc$^{-1}$).}
    \label{fig:NFW_fit}
\end{figure*}

If we now look at the other columns (blue, green and purple), we see the same key features described above, but to variable degrees. The inner density profile of the middle left panel (matched initial amplitude) is almost identical with the equivalently matched simulations in the left panel. Similarly, the splashback radii occur at roughly the same location but the effect is stronger in this case, with steeper minimum slopes. Looking at the middle right panel, which corresponds to variable $n_s$ but with a \textit{Planck} pivot that effectively decreases the variation in the initial amplitude of the power spectra at modes sampled in the box, we see that the qualitative dependence on $n_s$ is the same (more cuspy inner profiles with increased $n_s$, the splashback radius having a non-monotonic dependence on $n_s$, etc) but to a milder degree than seen in the left panel. The right panel, which corresponds to the $k_{pivot}=1 h$ Mpc$^{-1}$ suite all with similar initial amplitudes, shows that the density profiles are almost identical with indistinguishable inner slopes and only slight differences at larger radii. The general trend at high radii for this suite (purple) is for the higher $n_s$ values to have steeper slopes out to the splashback radius. 

With the middle left panel being broadly indistinguishable from the fiducial suite, the middle right panel exhibiting the same dependence on $n_s$ but to a more subdued level and the right panel being broadly independent of $n_s$, it can be concluded that the dominant effect is due to the change in amplitude of the initial power spectra at modes sampled within the box, as opposed to changes in the slope of the initial power spectra. This is not to say that changes in the shape have no effect whatsoever, as shown in other works \citep[e.g.][]{Dalal2010,Diemer_2015,Aaron_shape} and as we will discuss later in Section~\ref{mass_dependence_peak_height}.

One particularly interesting result seen here is the exact values of the logarithmic slope. At the outer radii, but within the splashback radius, we see in some cases the slope reaches particular steep values, with slopes of $\gamma_{\rho}(r) \approx -3.5$ in the most extreme cosmologies. By definition, the asymptotic outer slope of an NFW profile is $\gamma_{\rho}=-3$. This, therefore, indicates that these density profiles will not be well fit by an NFW profile. Indeed this is the case, as we show in Fig.~\ref{fig:NFW_fit}.\footnote{See Section~\ref{mass_dependence_peak_height} for details of how the best fit parameters are obtained, as well as the figure of merit used.} It is worth highlighting that this is in the radial regime that for `normal' cosmologies an NFW profile would be a good fit for stacked relaxed halos. Although these profiles cannot be fit with an NFW profile they can be well fit by an Einasto profile, see Fig.~\ref{fig:NFW_fit}.  Note that it is not surprising that an Einasto profile fits the profiles better, as it has an additional free parameter and, unlike the NFW profile, there is no finite limit on the inner or outer slope.

It is interesting to note that, if one \textit{assumes} an Einasto profile instead of an NFW form, then such steep logarithmic outer slopes are expected within $r_{200}$ for haloes that are sufficiently concentrated. The logarithmic slope of an Einasto profile can be written as $\gamma_{\rho} (r)=-2(cr/R_{\rm{200c}})^{\alpha}$. Hence, for a profile with $\alpha=0.16$ we would expect the slope of density profile to be steeper than $-3$ within $R_{\rm{200c}}$ if $c>12.6$. Additionally, for $\gamma_{\rho}(r=0.6R_{\rm{200c}})=-3.5$ (the most extreme case in the simulations), one requires $c \approx 50$, which is consistent with the concentrations we measure (see Section~\ref{mass_dependence_peak_height}).

\begin{figure*}
    \centering
    \includegraphics[width=2\columnwidth]{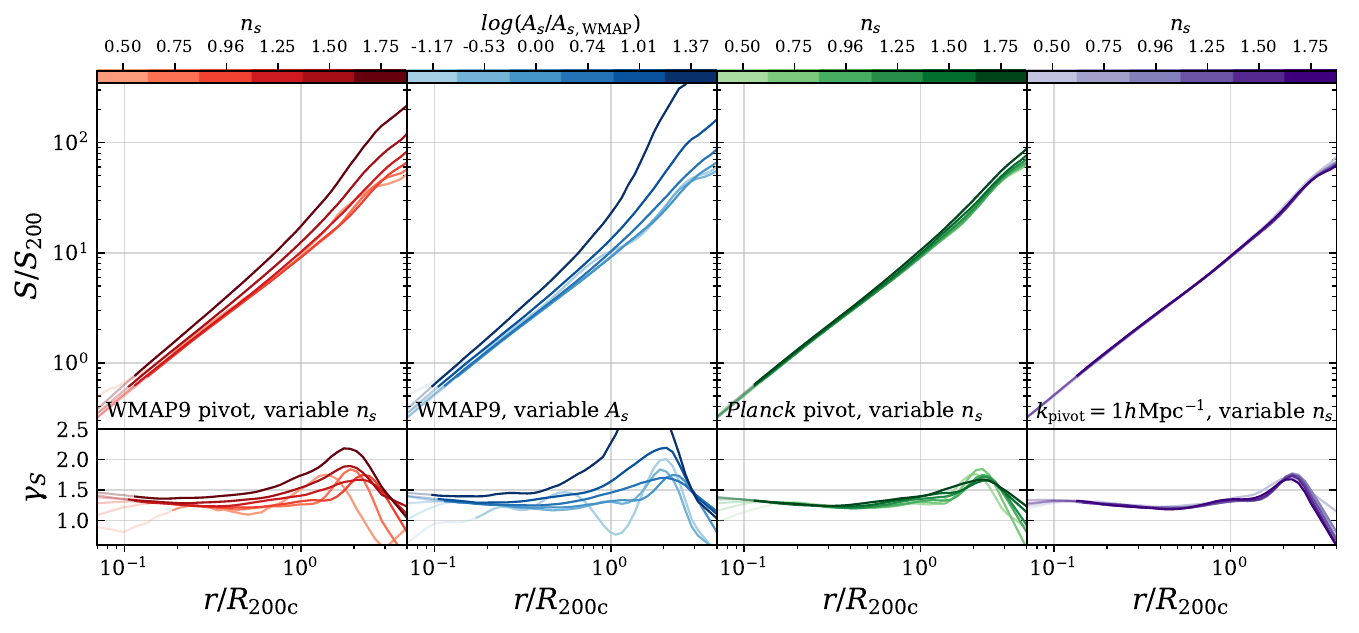}
    \caption{Entropy profiles of haloes in the mass range $M_{200c}=10^{13}$--$10^{13.5} h^{-1}$ M$_{\odot}$. In the top panels is plotted the entropy, defined as $S(r)=(\rho/\sigma^3)^{-2/3}$, normalised by a `virial' entropy, $S_{200c}=(200\rho_{\rm{crit}}/v_{\rm{circ}, 200c}^3)^{-2/3}$. The bottom panel shows the logarithmic slope of those profiles, $\gamma_{S}(r)=d(ln(S))/d(ln(r))$. See Fig.~\ref{fig:dens_profile} for a description of the general structure of the figure. In general it is observed that the slope of the entropy profiles are not constant and in general depend on the underlying cosmology. Similar to the density profiles, the dominant factor in determining the entropy profiles is the amplitude of the linear power spectra at an associated $k$-mode.}  
    \label{fig:entropy_profile}
\end{figure*}

It has been shown in other work that an Einasto profile is a better fit to the density profiles of DM haloes than and NFW, even with a fixed shaped parameter \citep{Nevaro2004,merritt2006,navarro2010, wang2019universality}. The key result here is not that the Einasto provides a better fit, but that the outer asymptotic slopes have $\gamma_{\rho}<-3$, which is incompatible with the NFW form. The strong deviations from the NFW form appears to be at odds with some literature that state it is independent of changes to the initial power spectrum \cite[e.g.][]{NFW_indepent_ICs1,NFW_indepent_ICs2,NFW_indepent_ICs3}. This statement is usually based upon work that has studied cosmologies with differently shaped power spectra to $\Lambda$CDM; such as a strict power law or that associated with hot and warm dark matter. The main differences between these works and our own appears to be the amplitudes of the initial power spectra. In our work the most extreme cases are seen in the simulations with the largest initial amplitudes, while other works have predominantly fixed the amplitude (usually through $\sigma_8$).

\subsection{Stacked entropy profiles}

In Fig.~\ref{fig:entropy_profile} we show the stacked entropy profiles of the various suites of simulations. Haloes are stacked by taking the median entropy of individual haloes in units of $r/R_{\rm{200c}}$. The colour scheme and ordering of the panels in the plot are identical to that of Fig.~\ref{fig:dens_profile}.

It is apparent that the entropy profiles in Fig.~\ref{fig:entropy_profile} are much more similar to one another than the density profiles and that these profiles are much closer to being scale independent. Let us focus initially on the fiducial suite (left panels). As noted previously in the literature the entropy profiles appear to have a simple power law form at smaller radii. It is clear that the inner logarithmic slope is dependent on $n_s$, highlighting that the slope of the power low does not appear to be a constant in this suite. As we discuss later, in general the slope of the entropy profiles exhibit a clear mass dependence and are \textit{not} a universal constant. It also appears that for the individual simulations the inner logarithmic slopes are not exactly constant with slight radial dependences, as such the entropy profiles are not a perfect power law.

At small radii the slopes are approximately constant but a sharp spike is observed at larger radii, this feature corresponds to the splashback radius. The splashback radius is therefore not only a feature in the density of a halo but also in its entropy (or, equivalently, its phase-space density). The entropy profiles of DM therefore exhibit very similar behaviour as the true entropy of gas in galaxy clusters: approximately power law behaviour within $R_{\rm{200c}}$ and a sharp increase at high radii. In clusters this feature in the gaseous entropy corresponds to the outer shock radius and not directly to the splashback radius (as the DM entropy does), however, the two are strongly linked as discussed in \citet{Lau15}. Studying the density and velocity dispersion profiles individually the increase in entropy gradient at the splashback radius is predominantly due to the change in the gradient of the density profile, as opposed a change in the gradient of the velocity dispersion profile.

Interestingly, it can be seen in the logarithmic slope that the simple power law nature of the entropy profiles changes somewhat before the splashback radius, as $\gamma_{S}$ increases noticeably before the spike corresponding to the splashback radius. This behaviour has been noted before in the literature (see for instance \citealt{Ludlow_ppsd}) and is actually predicted by secondary infall models \citep{Bertschinger}. In these models this break in the power law at higher radii is caused by outer mass shells that have yet to be fully virialised and reach stable orbits. 

When looking at the other sets of simulations (blue, green and purple lines) we see the same qualitative behaviour as for the density profiles. The variable $n_s$ with WMAP pivot (left) and matched $A_s$ (middle-left) behave almost identically, the \textit{Planck} pivot (middle-right) has the same qualitative behaviour as the other two but to a milder degree, and the $k_{\rm{pivot}}=1 h$ Mpc$^{-1}$ (right) are almost all identical with indistinguishable inner behaviour but with slightly different outer profiles. 

It appears that the inner slope of the entropy profiles are, in general, dependent on changes to the initial power spectra. The inner slopes vary in the range $\gamma_{S} \approx 1.2$--$ 1.4$ with most simulations exhibiting a slight increase in logarithmic slope to lower radii. These values are within the range quoted in the literature \citep[e.g.][]{taylor-navarro, ppsd_1.94, faltenbacher_entropy}. Our results therefore suggest that the entropy profiles are, in general, not precisely simple power laws within the splashback radius. Instead the logarithmic slopes have a slight radial dependence. This appears to be consistent with claims from previous work (e.g., \citealt{ppsd_not_pow}) that the PPSD profiles have slopes that vary mildly with radius over the range sampled by simulations. 

Although a simple power law does not provide a perfect description of the shape of the entropy profiles, it does provide a reasonably good approximation over the ranges that sampled. As such, we fit the entropy profiles with simple power laws to look at the dependence on mass and other factors, in Section~\ref{mass_dependence_peak_height}.

\begin{figure*}
    \centering
    \includegraphics[width=2\columnwidth]{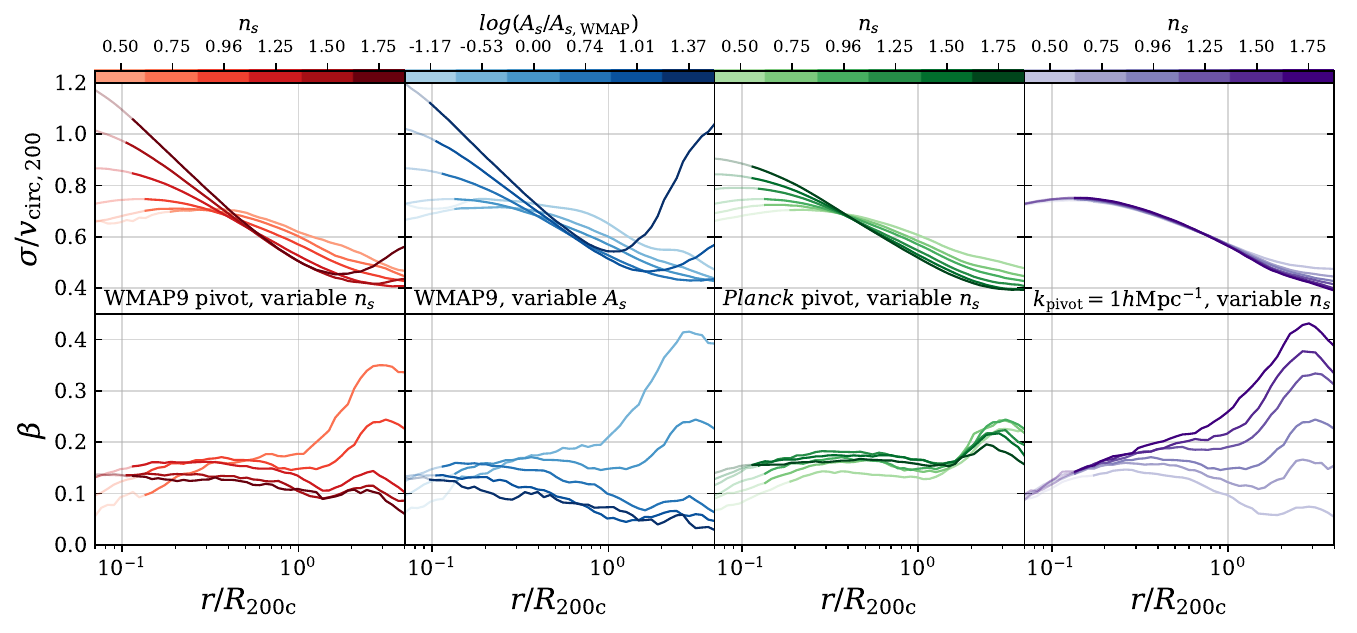}
    \caption{Stacked velocity dispersion and velocity anisotropy profiles of haloes in the mass range $M_{200c}=10^{13}$--$10^{13.5} h^{-1}$ M$_{\odot}$. In the top panels is plotted the total velocity dispersion while the bottom panel shows the velocity anisotropy, $\beta=1-\sigma_{T}^2/\sigma_{r}^2$. See Fig.~\ref{fig:dens_profile} for a description of the general structure of the figure.}
    \label{fig:vel_profile}
\end{figure*}

\subsection{Stacked velocity dispersion profiles}

In Fig.~\ref{fig:vel_profile} we present the stacked total velocity dispersion, $\sigma$ and velocity anisotropy profiles, $\beta$, as a function of radius, see Section.~\ref{density and velocity calculation} for definitions. The structure of the plot and which colours correspond to which simulation are identical to that of Fig.~\ref{fig:dens_profile}.

We will again focus our attention on the fiducial suite of simulations first (left panels in red). Looking at the total velocity dispersions (top panel) we can see that larger values of $n_s$ result in velocity dispersion profiles that have larger inner velocity dispersions and smaller values at large radii (and vice-versa for smaller $n_s$). This is in qualitative agreement with what we would expect from equilibrium arguments such as Jeans theory, in order to support the changes in the mass structure discussed in Section \ref{density_section}. Interestingly, there does not appear to be any obvious feature in the velocity dispersion profiles corresponding to the splashback radius. Although not shown here, such a feature can be seen when looking at the logarithmic slope of the \textit{radial} velocity dispersion. That this feature is most obvious in the logarithmic slope of the profile agrees with the observed behaviour of the density distribution. The feature only being dominant in the radial velocity dispersion and not the total velocity dispersion is similarly expected; the splashback radius represents the first apocenter when the particles turn around we would expect little change to the angular velocities but a significant effect on the radial component.

The velocity dispersion profiles for the rest of the simulations (middle-left, middle-right and right) exhibit the same general trends seen in the density profiles (see Fig.~\ref{fig:dens_profile}). The fiducial and matched amplitude suites behave almost the same while the suite that uses a \textit{Planck} pivot point exhibit the same trends to a milder degree and the $k_{\rm{pivot}}=1 h$ Mpc$^{-1}$ suite are almost indistinguishable in the inner radii with only slight differences beyond $R_{\rm{200c}}$.

Looking at the velocity anisotropy profiles (bottom) the differences between the simulations is not as clearly pronounced as the velocity dispersion. The approximate inner values, $\beta \sim 0.1$--$0.2$, are consistent with that found in previous work \citep[e.g.][]{Ludlow_ppsd}. The inner behaviour of $\beta$ is approximately constant with larger values of $n_s$ leading to more isotropic particle orbits, except for $n_s=0.75$ that reverses this trend. At large radii, the velocity anisotropy increases strongly as we transition into a regime where matter is being actively accreted onto the halo. The small inner differences in $\beta$ suggest that haloes in the different simulations are roughly in the same dynamical state but with different density profiles. However, the more significant differences at large radii strongly suggest that they are accreting matter in different ways. It is also clear from the results of the $k_{\rm{pivot}}=1 h$ Mpc$^{-1}$ suite shown in Fig.~\ref{fig:dens_profile} and Fig.~\ref{fig:vel_profile} that the anisotropy-density slope relation is not universal as claimed by some works \citep[e.g.][]{Hansen2006}.

The velocity anisotropy profiles between the fiducial model and matched amplitude behave very similarly to the rest studied previously; broadly agreeing with the same amplitude differences but slightly different radial dependencies. The two suites with smaller amplitudes (green and purple) have much clearer radial profiles with a weakly parabolic-shaped curve as opposed to the roughly constant profiles seen for the two other suites. Interestingly, the suite of simulations with the most similar $\beta$ profiles is that which adopts a \textit{Planck} pivot instead of the $k_{\rm{pivot}}=1 h$ Mpc$^{-1}$ suite, the opposite trend of what was seen for the density, entropy and velocity dispersions profiles. It seems as though the inner behaviour of the $k_{\rm{pivot}}=1 h$ Mpc$^{-1}$ simulations is very similar but the outer dependence, but within the virial radius, on $n_s$ is drastically different, with the main change arising from the radius where $\beta$ begins to quickly increase. 

This highlights that, even though the density and velocity dispersion profiles are almost identical between the different values of $n_s$, the haloes are in different dynamical states and growing in distinctly different ways. The suite that adopts a \textit{Planck} pivot (middle-right panel), on the other hand, has clearly different inner $\beta$ values but a very similar outer behaviour.

There are two interesting features in all $\beta$ profiles occurring. The first feature occurs at roughly $R_{\rm{200c}}$, which is most clearly seen in the \textit{Planck} pivot and $k_{\rm{pivot}}=1 h$ Mpc$^{-1}$ suites (right two panels). The feature correspond to where $\beta$ suddenly increases. This feature likely demarcates an inner region of the halo that is in equilibrium from the region where matter is still being actively accreted. It is interesting that this radius roughly corresponds to $R_{\rm{200c}}$ but not precisely, with the $k_{\rm{pivot}}=1 h$ Mpc$^{-1}$ pivot case (right) having profiles that under- and over-shoot $R_{\rm{200}}$. This suggests that $R_{\rm{200}}$ gives a reasonable approximation to where the halo is in equilibrium, but it is not a perfect prediction. The second feature occurs at approximately $2R_{\rm{200c}}$ where the $\beta$ exhibit a local maxima, it is unclear what exactly this feature corresponds to but is likely related to the splashback feature observed in the density profiles.

\section{Mass and peak height dependence} \label{mass_dependence_peak_height}

\begin{figure*}
    \centering
    \includegraphics[width=2\columnwidth]{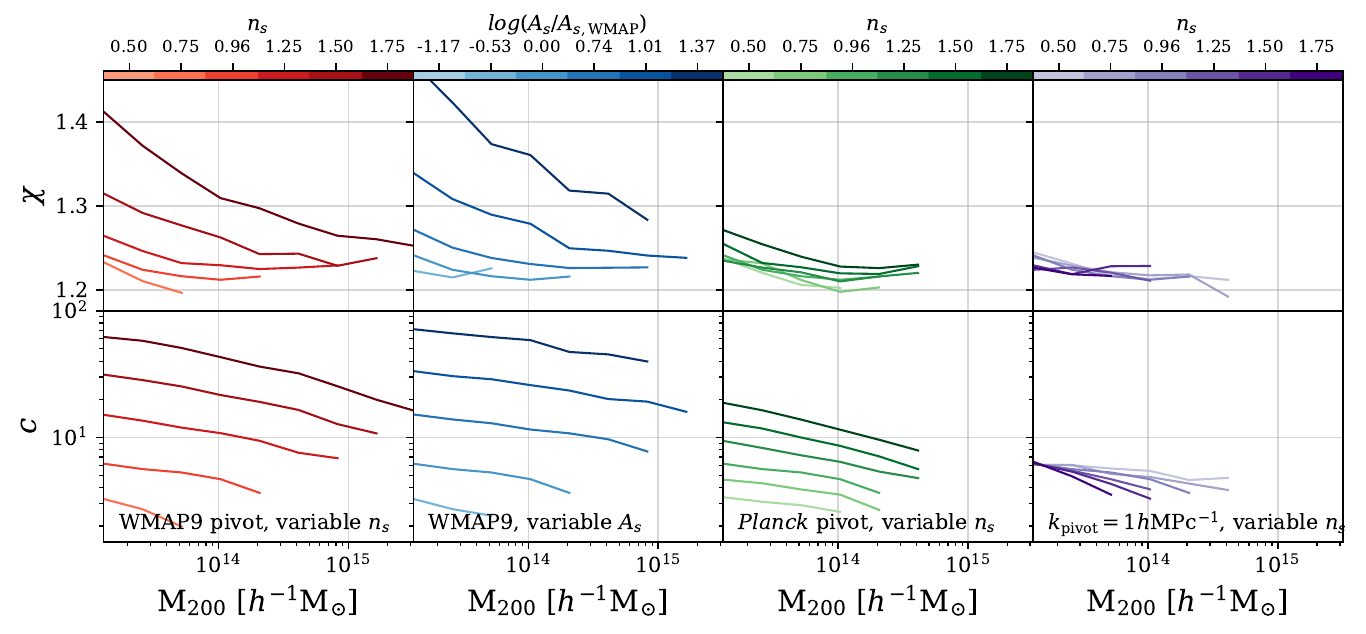}
    \caption{Fitted parameters as a function of mass. Haloes with greater than 2000 particles are stacked in mass bins of 0.3 dex, only bins with more than 15 haloes are plotted. The resulting stacked density and entropy profiles are then fit by minimising $\psi^2$, see Eqn.~(\ref{error}). There are two fitted parameters shown, halo concentrations, $c$, and entropy power law exponent $\chi$. Each row shows one of these parameters as a function of halo mass, $M_{\rm{200c}}$, while each column represents the four suites of simulations; WMAP pivot variable $n_s$ (left), matched $A_s$ (middle-left), \textit{Planck} pivot point with variable $n_s$ (middle-right) and $k_{\rm{pivot}}=1h^{-1}$ M$_{\odot}$ variable $n_s$ (right). The different shades represent the particular value of $n_s$, or $A_s$, see colourbar.}
    \label{fig:mass_parameters}
\end{figure*}

We have demonstrated that varying the initial power spectrum can have a significant effect on the density and entropy profiles of collapsed structures. Here we study how these profiles vary with halo mass and peak height.

The following figure of merit is minimised to obtain the best fit parameters for a given fitting routine and data set,
\begin{equation}
    \psi^2=\sum_i [log(y_i)-log(y_{\rm{fit}})]^2.
\end{equation}
Where $y_i$ is the set of data (either density of entropy in this work) and $y_{\rm{fit}}$ the prediction from the specific profile used to fit.

Halo density profiles are fit with an Einasto curve \citep{einasto},
\begin{equation} \label{einasto}
    \rho_{\rm{E}}(r)=\rho_{-2}exp \bigg(-\frac{2}{\alpha} \bigg[\bigg(\frac{r}{r_{-2}} \bigg)^\alpha-1 \bigg] \bigg).
\end{equation}
where $r_{-2}$ is the radius where the logarithmic density slope is $-2$, $\alpha$ is a shape parameter that quantifies how quickly the density varies as a function of radius. The normalisation, $\rho_{-2}$, is equal to the density at $r_{-2}$. Although we use the above equation to fit the profiles, we do not quote $r_{-2}$ values but instead concentration, defined as $c = R_{\rm{200c}}/r_{-2}$. We use a fixed $\alpha=0.16$ when fitting these profiles as we find that when left to vary it clusters around this value but with significant scatter, it is also consistent with values quoted in the literature \citep[e.g.][]{Gao_alpha_mass}. Additionally, it is found that $r_{-2}$ and $\alpha$ are strongly correlated, fixing $\alpha$ allows us to significantly reduce the noise while providing approximately the same values for $r_{-2}$. The scale radius, $r_{-2}$ is not always resolved in our simulations, as can be seen in Fig.~\ref{fig:dens_profile}. Note that the inferred values of $r_{-2}$, and similarly $c$, presented in this work represent the best-fit values over the resolved radial range, whether $r_{-2}$ is directly sampled or not.

To fit entropy profiles we use a power law,
\begin{equation} \label{eqn:entropy_fit}
    S(r)/S_{\rm{200c}}=\Delta_S(r/R_{\rm{200c}})^{\chi}.
\end{equation}
where $\Delta_S$ is a normalisation parameter, equal to the virial normalised entropy ($S/S_{\rm{200c}}$) at $R_{\rm{200c}}$, and $\chi$ is the exponent of the power law.

\begin{figure}
    \centering
    \includegraphics[width=\columnwidth]{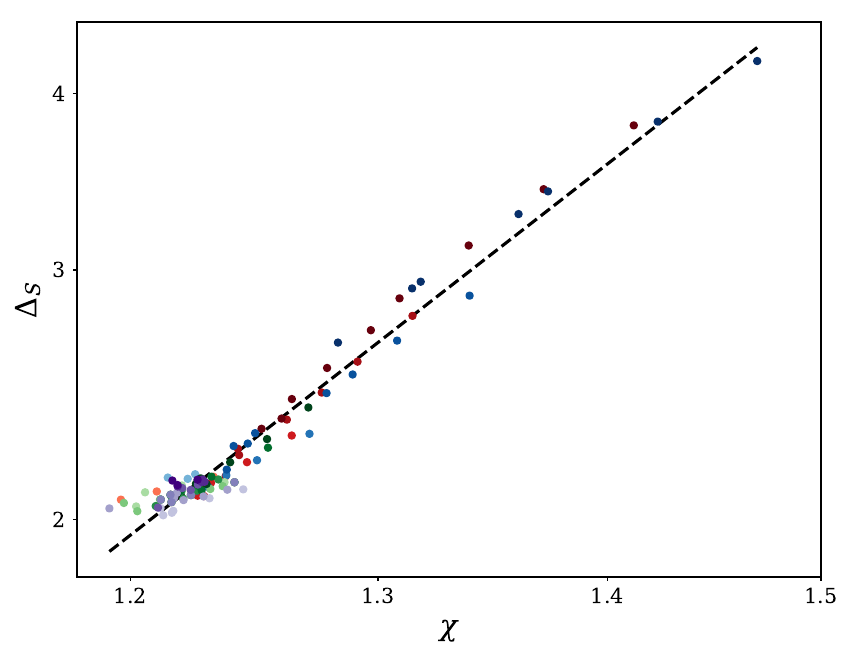}
    \caption{Entropy normalisation, $\Delta_S$, as a function of power law exponent, $\chi$, see Eqn.~\ref{eqn:entropy_fit} for definition, for all cosmologies sampled. Each data point represents a mass bin with the colour denoting the particular cosmology, as originally specified in Fig.~\ref{fig:initial_pow}. The black dashed line represents the best fit power law relation of $\Delta_S = 0.94 \chi^{3.92}$. There is no discernible dependence on the initial power spectra and the relation appears to be universal.}
    \label{fig:chi_delta_s}
\end{figure}

\subsection{Mass dependence}

In Fig.~\ref{fig:mass_parameters} we present the dependence of $c$, $\chi$ and $\Delta_S$ on halo mass, $M_{\rm{200c}}$. We present here the normalisation of the entropy but not the density, $\rho_{-2}$, as it can be found through the overdensity definition of $M_{\rm{200c}}$ and $R_{\rm{200c}}$, however $\Delta_S$ cannot. Each row shows a different parameter's dependence on mass, while each column represents the different set of simulations with colours consistent with the rest of the paper. Note the left and middle-left panels do not have data plotted for the $n_s=0.5$ case as there are too few haloes above the mass cut to examine the mass dependencies. 

As before we will focus first on the fiducial suite, shown in the left column of Fig.~\ref{fig:mass_parameters}. Looking at the concentration, we see the same bulk trends as found for the stacked density profiles (see Fig.~\ref{fig:dens_profile}); larger values of $n_s$, corresponding to larger amplitudes in this suite, results in haloes that are more concentrated, with values as high as $c \approx 70$. For all values of $n_s$, the concentration monotonically decreases with mass and in general the concentration--mass relations for the different cosmologies have similar slopes but distinctly different amplitudes.

We now turn our attention to the mass dependence of the entropy profiles in the various simulations. Firstly we notice that, in general, the slope, $\chi$, does depend on halo mass. However, for the standard $n_s=0.96$ case, the dependence on mass is quite mild with an almost constant value of $\chi \approx 1.21$, in agreement with previous studies on the PPSD profile (see the Introduction). For all simulations $\chi$ monotonically decreases with mass, with larger $n_s$ resulting in a stronger mass dependence and values of $\chi$ at the low mass end distinctly larger than $1.25$ in the more extreme cosmologies. The fact that the slope and its dependence on halo mass varies with $n_s$ in these suites demonstrates that there is nothing particularly special about the $\chi\approx1.25$ result from the standard cosmology (i.e., with $n_s=0.96$), calling into question the perceived fundamental nature of the, previously, apparent universality of the PPSD. 

As mentioned previously $\Delta_S$, the amplitude of the normalised entropy profiles, is not specified by the overdensity definition of halo mass and radius, i.e. $R_{\rm{200c}}$ and $M_{\rm{200c}}$, unlike the density normalisation. As such we have additionally studied how $\Delta_S$ varies in these cosmologies. Recall that $\Delta_S$ is defined such that it is the amplitude of the virial normalised entropy at $R_{\rm{200c}}$, see Eqn.~\ref{eqn:entropy_fit}. From Fig.~\ref{fig:entropy_profile} it is clear that for haloes with a larger slope, $\chi$, correspond to larger values of $\Delta_S$. Indeed, this is a general feature of all of the cosmologies studies here with there appearing to be a universal relation between the entropy slope and normalisation that is independent of primordial amplitude or spectral index, as shown in Fig.~\ref{fig:chi_delta_s}. The relation between $\chi$ and $\Delta_S$ is well fit by 

\begin{equation}
    \Delta_S = 0.94 \chi^{3.92}.
\end{equation}

Looking at all sets of simulation we can see that the dominant behaviour of $c$ and $\chi$ is due to the amplitude change of the primordial power spectra on scales sampled in the box, as opposed to the shape ($n_s$) change of the initial power spectra. The fiducial model (left most panel) and matched amplitude (middle-left) have the closest $c$ and $\chi$ values for equivalent masses, while the \textit{Planck} pivot suite (middle-right) show the same general trends but to a milder degree. The $k_{\rm{pivot}}=1 h$ Mpc$^{-1}$ suite (right most panel) has the most similar dependence on mass between different values of $n_s$, with roughly the same amplitude but distinctly different detailed dependencies. Although the bulk of the changes in the trends appear to be due to the amplitude change in the primordial power spectrum, the fiducial and matched $A_s$ suites do have different mass dependencies. In general it seems that larger initial amplitudes lead to larger values of $c$ and $\chi$; for $n_s>0.96$ the matched amplitude simulations have more power for smaller $k$ corresponding to larger massed objects, and vice-versa for $n_s<0.96$. The results here are then consistent with the bulk offsets, at high masses the matched $A_s$ simulations have comparably larger $c$ and $\chi$ values than the fiducial suite.

The concentrations in the $k_{\rm{pivot}}=1 h$ Mpc$^{-1}$ suite exhibit a particularly notable feature that they all converge at $M_{\rm{200c}} \approx 10^{13} h^{-1} \text{M}_{\odot}$, see bottom right panel of Fig.~\ref{fig:mass_parameters}. This is, in part, by construction. The pivot point was specifically chosen to correspond to haloes of approximately this mass. The interpretation is therefore that the dominant factor in setting the concentration of a halo is the amplitude of the associated k-mode. It is plausible that the deviations above $M_{\rm{200c}} \approx 10^{13} h^{-1} \text{M}_{\odot}$ and different dependences on mass are predominately due to the amplitude of the power spectra being different in $k>1 h$ Mpc$^{-1}$ (see Fig.~\ref{fig:initial_pow}) as opposed to differences in the slope directly. 

\subsection{Peak height dependence} \label{Peak_height_dependence}

We now turn our attention to peak height, which is defined as
\begin{equation} \label{peak_height_1}
    \nu = \frac{\delta_c}{\sigma(M,z)}.
\end{equation}
\noindent where $\delta_c$ is the condition for collapse\footnote{We take $\delta_c=1.686$ and ignore the mild redshift and cosmology dependence.} in the spherical collapse model \citep{Gunn_&_Gott} and $\sigma$ is the RMS of density fluctuations. $\sigma(M,z)$ itself is calculated from the linear power spectra, $P(k)$, by
\begin{equation} \label{peak_height_2}
    \sigma ^2(R)=\frac{1}{2 \pi^2} \int^{\infty}_{0}k^2P(k) |\tilde{W} (kR)|^2 dk.
\end{equation}
$\tilde{W} (kR)$ is the Fourier transform of a spherical top hat function. Halo mass is converted to a radius by projecting onto a sphere with the same average density as the universe, i.e. $R^3=M/(4/3 \pi \rho_{\rm{m}})$, and then peak height calculated using Eqn.~(\ref{peak_height_1}). The intuitive interpretation of peak height is to quantify, on average, how old and how rare a halo is, with larger values of $\nu$ corresponding to rarer, younger objects.

It is well established that for a given cosmology the $c$--$M$ relation varies with redshift and when comparing different cosmologies this behaviour is even more complex. However, previous work has shown that the $c$--$M$--$z$ relation can approximately be reduced to a peak height--concentration relation, effectively removing the redshift dependence \citep[e.g.][]{prada2012, Ludlow_CDM}. Although it is clear that peak height is the dominant factor in setting the halo concentration there are complex secondary terms required to accurately predict the concentration at all redshifts for a given cosmology. Halo concentration is expected to scale much more closely with peak height than mass for two main reasons; (i) in calculating peak height any change to the cosmology is implicitly accounted for by integrating over P(k) (see Eqn.~(\ref{peak_height_2})); and (ii) the formation time and age of a halo, which is intimately linked to the concentration \citep[e.g.][]{bullock2001, Wechsler_2002, zhao2003}, is expected to scale with peak height and not, in general, with mass. Although the dominant factor in setting the density profiles of a halo is indeed the peak height, there is a residual dependence on the shape of the power spectrum as shown in \citet{Diemer_2015}, leading us to study the dependence on $n_s$ directly. 

\begin{figure*}
    \centering
    \includegraphics[width=2\columnwidth]{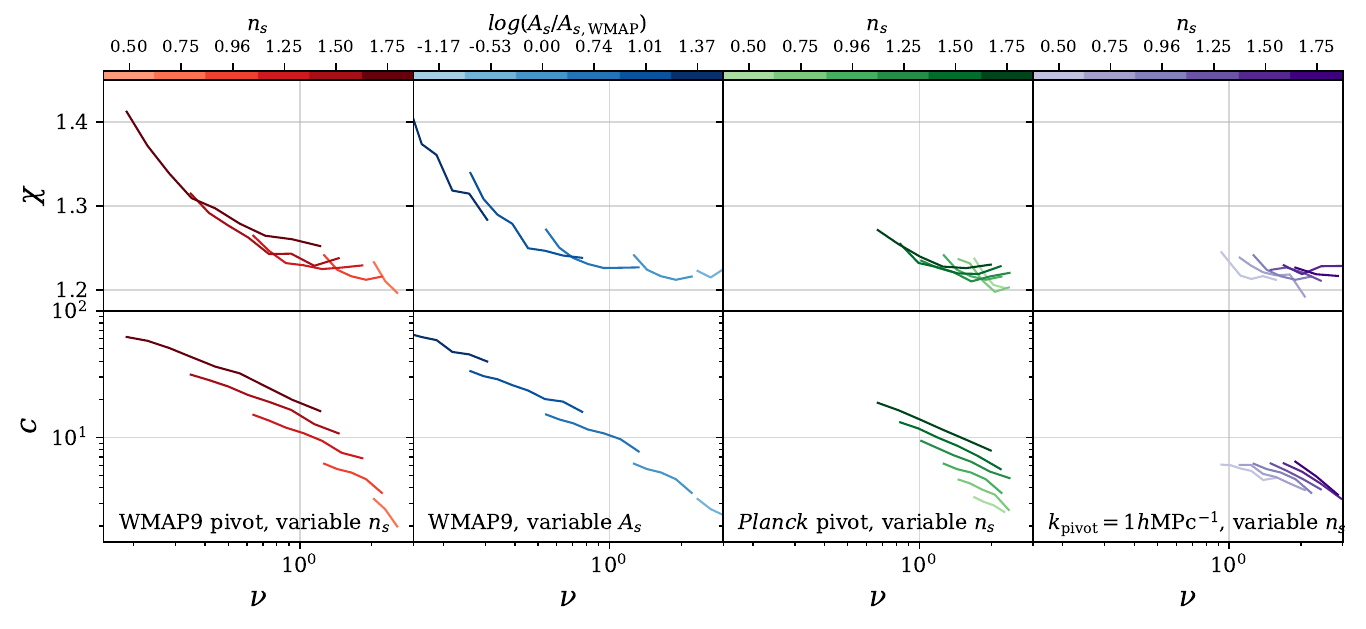}
    \caption{Fitted parameters, halo concentration, $c$, and entropy power law exponent, $\chi$, as a function of peak height, $\nu$. See Eqn.(\ref{peak_height_1}) for the definition of $\nu$. The plot, and associated analysis, is identical to Fig.~\ref{fig:mass_parameters} but with the mass converted to peak height for each different cosmology.  In general amplitude changes to the primordial power spectra are well described by peak height, however there are secondary `shape' changes to the linear power spectra that are not accounted for by a change in peak height.}
    \label{fig:overdensity_parameters}
\end{figure*}

In Fig.~\ref{fig:overdensity_parameters} we present how the fitted density and entropy parameters, $c$ and $\chi$, vary with peak height, $\nu$, at $z=0$. The structure of the plot is identical to that of Fig.~\ref{fig:mass_parameters}. Let us first focus on how these parameters vary for the two suites that are dominated by amplitude changes to the initial power spectrum (left and left-middle panels). We see that for these suites all of the fitted parameters are close to a single function of $\nu$, with the large amplitude offsets seen when plotted as a function of mass removed. The $c$--$\nu$  relation appears to be closest to a single function in the suite with fixed $n_s$ and matched $A_s$, however all simulations exhibit slight but clearly different dependences on $\nu$.

It is clear in all suites that peak height is the dominant term in setting $\chi$ and $c$, however, there is a secondary dependence observed for both changes to $A_s$ and $n_s$. Focusing on the parameters in the suites dominated by a shape change to the power spectra (right-middle and right panels) we see that they are very clearly not described by a single function of $\nu$, with the $k_{\rm{pivot}}=1~h$ Mpc$^{-1}$ suite becoming even more stratified than when plotted as a function of mass. This stratification and strong dependence on $n_s$, as well as $\nu$, is most clearly seen in the concentrations. We therefore conclude that, in general, the concentration--peak height relation is a function of $A_s$ and $n_s$, but is more more sensitive to changes in the shape, or slope, of the power spectra than amplitude differences. The result here that the $c-\nu$ relation depends strongly on the shape of the power spectra is in qualitative agreement with the works of \citet{Diemer_2015} who showed that for scale free cosmologies concentration has a different dependence on peak height for different slopes of the linear power spectra. Although we have not studied how $\alpha$, the shape parameter in the Einasto profile, varies with mass or redshift in these cosmologies we would expect the same general behaviour that we see for concentration. \citet{Ludlow2017} showed that, again, for a scale free universe the $\alpha \text{--} \nu$ relation is redshift independent for a given cosmology but does depend on the slope of the linear power spectrum.

\begin{figure*}
    \centering
    \includegraphics[width=2\columnwidth]{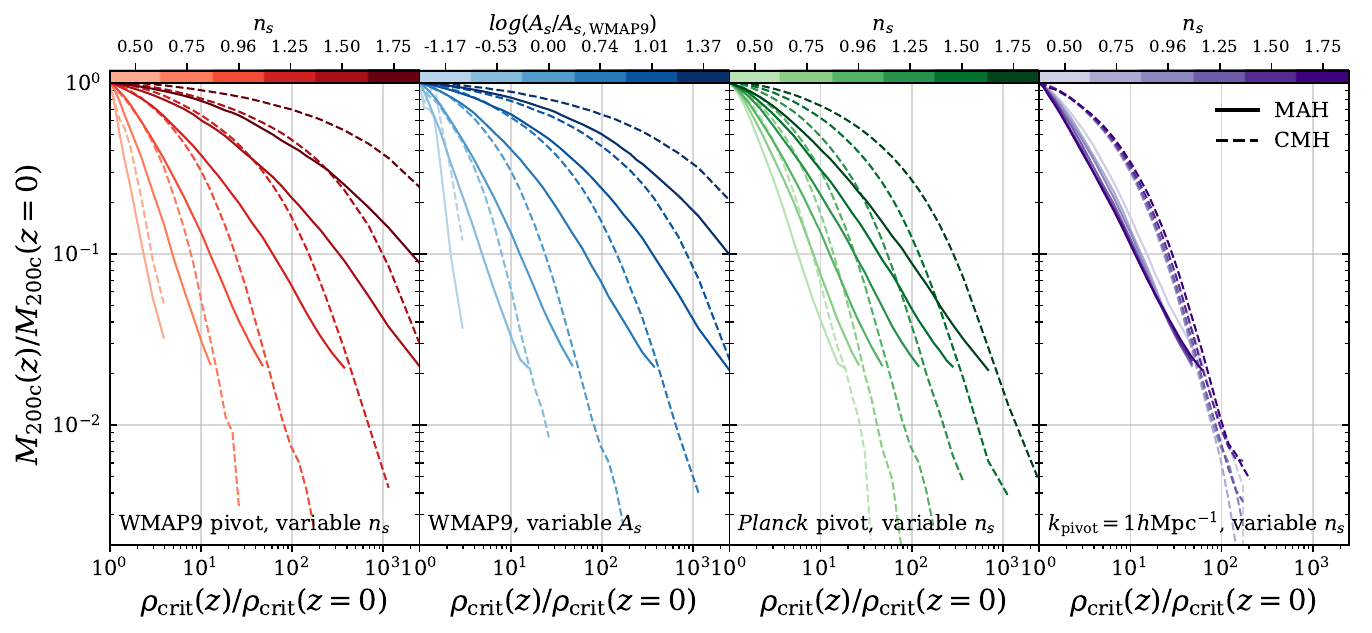}
    \caption{The normalised MAH and CMH plotted as a function of critical density, for stacked haloes in the mass range $M_{200c}=10^{13}$--$10^{13.5} h^{-1}$ M$_{\odot}$. See \ref{mah_cmh_subsection} for the definition of MAH and CMH. The different panels and colours represent the different suite of simulations, see label in bottom left of each panel. The shades of each colour represent the value of $n_s$, or matched $A_s$, used for the simulations, see colourbar. The different line-styles represent either the MAH or CMH, see legend. The changes to the MAH and CMH of haloes is in qualitative agreement with what is expected from the change in their concentration. Haloes with higher (lower) concentrations are those that formed earlier (later) with subsequent slower (faster) accretion today.}
    \label{fig:MAH}
\end{figure*}

\subsection{Interpreting \texorpdfstring{$A_s$}{As} as a change of redshift}

The results observed in the suite with only an amplitude change to the primordial power spectra are qualitatively similar to what is seen for a single cosmology at multiple redshifts. This is not surprising as the the difference in the linear power spectra between different redshifts is purely an amplitude one. This is often expressed through the linear growth factor, $D(z)$,
\begin{equation}
\label{Eqn: Matched_amplitude}
    P(k,z)=D^2(z)P(k,z=0).
\end{equation}
Therefore cosmologies with different values of $A_s$ will have identical linear power spectra at different redshifts. This suggests that the effects of changes to the amplitude seen in Fig~\ref{fig:mass_parameters} \& \ref{fig:overdensity_parameters} can be expressed as a change in redshift. Quantitatively remapping changes in $A_s$ to a change in redshift would involve the detailed modelling of the secondary effects on concentration mentioned previously, which is beyond the scope of this work. However, we can still discuss the qualitative expectations. In general we would expect universes with $A_s<A_{s,\rm{WMAP}}$ at $z=0$ to look very similar to our own Universe at higher redshift. Similarly we would expect cosmologies with $A_s>A_{s,\rm{WMAP}}$ at $z=0$ to be like our own Universe in the future. Even though this is in general true it is found that the cosmologies with $A_s>A_{s,\rm{WMAP}}$ sampled in this work are actually more extreme than our own universe will ever be. This is due to the accelerated expansion, caused by the cosmological constant, suppressing the growth of structure to such a degree that our own Universe will never reach as large amplitudes as sampled by some of the cosmologies in this work (i.e. $D^2_{\rm{max}}<A_s/A_{s,\rm{WMAP}}$).

\section{Accretion histories and (semi-)analytic models} \label{MAH_CMH}

It is generally thought that concentration is predominantly determined by the formation time of a halo \citep[e.g., ][]{eke2001,bullock2001,Wechsler_2002}. Halo formation is sometimes described as an inside out process, \citep[e.g.][]{NFW_indepent_ICs2,zhao2003,Hoffman_2007,Ludlow13}, with a collapsed bound core forming early on in the universe with later accretion not penetrating this inner core and instead relaxing to larger radii. Under this interpretation, haloes that form earlier will be more concentrated. As, on average, low-mass haloes form earlier than high-mass haloes, this mechanism naturally accounts for the qualitative trend of the $c$--$M$ relation in a standard cosmology.

In this section we will check whether this interpretation of the $c$--$M$ relation agrees with our results for more extreme variations of the initial conditions by directly comparing the accretion histories and formation times of haloes in our simulations. We also compare the predictions of halo concentration from a range of recent models \citep{correa_3,Aaron_cm_relation_warm,Diemer_2019} to the results seen in this work.

\subsection{Mass accretion and collapsed mass histories} \label{mah_cmh_subsection}
We study here if the link between halo concentration and mass accretion history (MAH), and similarly the collapsed mass history (CMH) (both quantities are defined below), still holds in these extreme cosmologies. 

The MAH and the CMH are two separate but complimentary statistics that reduce the full merger history of haloes into more tangible quantities. The MAH is defined by following the most massive progenitor at each snapshot; how the mass, $M_{\rm{200c}}$, of these progenitors varies with redshift is then the MAH. The MAH is a useful statistic for the growth of a halo, highlighting how it's most massive branch of its merger tree grows with redshift. It does not, however, encapsulate the full spectrum of accreted collapsed structure, particularly ignoring smaller haloes that contributed to the final halo. The CMH, on the other hand, better encompasses the full plethora of accreted structure. The CMH is defined as all progenitors that have collapsed by redshift, $z$, above a certain fraction, $f$, of the final mass. In this work we use $f=0.02$, but this is in principle a free parameter in defining the CMH (see \citet{Aaron_cm_relation_warm} for a discussion of how the CMH changes with $f$). To calculate the stacked MAH we take the median from all haloes in the quoted mass range, while for the CMH we take the mean. All merger trees are generated using the algorithm presented in \citet{Merger_tree_algorithm}, using the $M_{\rm{200,c}}$ of the central halo to keep the mass definition consistent.

In Fig.~\ref{fig:MAH} we present the MAH and CMH histories for the various suites of simulations as a function of critical density, the mass range sampled is the same as that of Fig.~\ref{fig:dens_profile}--\ref{fig:vel_profile} so is directly comparable to them. The colours are the same as previous figures, with the line style denoting either the MAH or CMH (see legend). Initially focusing on the fiducial suite, left panel (red), the results are what is expected from our current understanding; the most cuspy haloes correspond to those that formed earliest. The general trend is for larger values of $n_s$ to lead to haloes that collapsed earlier, these haloes then accrete most rapidly in the earlier universe before growing more slowly today, and vice versa. This difference in growth is clearly dominated by the amplitude change to the initial power spectra as opposed to the shape change of this suite. The fact that the largest values of $n_s$ are accreting the slowest today implies they are more relaxed haloes, highlighting that haloes with density profiles most different from NFW and with the steepest outer slopes are actually those which are most relaxed. Looking at the other suites we again see changes in qualitative agreement with the differences in the density profiles: the matched amplitude simulations, middle left (blue), exhibit very similar changes to the MAH and CMH compared to the fiducial suite with the same bulk offset but slight different redshift dependences, the suite with \textit{Planck} pivot point, middle right (green) exhibits the same general trends with $n_s$ but to varying degree, while the $k_{\rm{pivot}}=1 h$ Mpc$^{-1}$ suite exhibit almost the same MAH and CMH. Our results appear to be consistent with the picture proposed in \citet{Dalal2010}. While the dominant term appears to be the amplitude of initial power spectra, i.e. peak height, the initial collapse and subsequent halo concentration is sensitive to the shape of the power spectra, hence the $c \text{--} \nu$ relation not being universal.

It seems that the observed trends in the density profiles discussed in Section \ref{density_section} are consistent with the current consensus as to their origin, with haloes more concentrated forming earlier when the universe was much denser. The key qualitative changes between the MAH/CMH match well those of the density profile, however it is unclear if this is in quantitative agreement of current models. We discuss this next.

\subsection{Comparison to analytic models}

In this section we compare our results to three notable works that aim to predict halo concentrations for a general cosmology. Specifically the works of \citet{correa_3}, \citet{Aaron_cm_relation_warm} and \citet{Diemer_2019}. The predictions from \citet{Aaron_cm_relation_warm} and \citet{Diemer_2019} are generated using the publicly available python package \texttt{COLOSSUS} \citep{colossus}, while the predictions \citet{correa_3} are generated uses the publicly available \texttt{COMMAH} package. In Fig.\ref{fig:CM_prediction} we compare the predictions of the aforementioned models to our results. Below we discuss each model in turn.

\subsubsection{Comparison to the model of Correa et al.}
\begin{figure*}
    \centering
    \includegraphics[width=2\columnwidth]{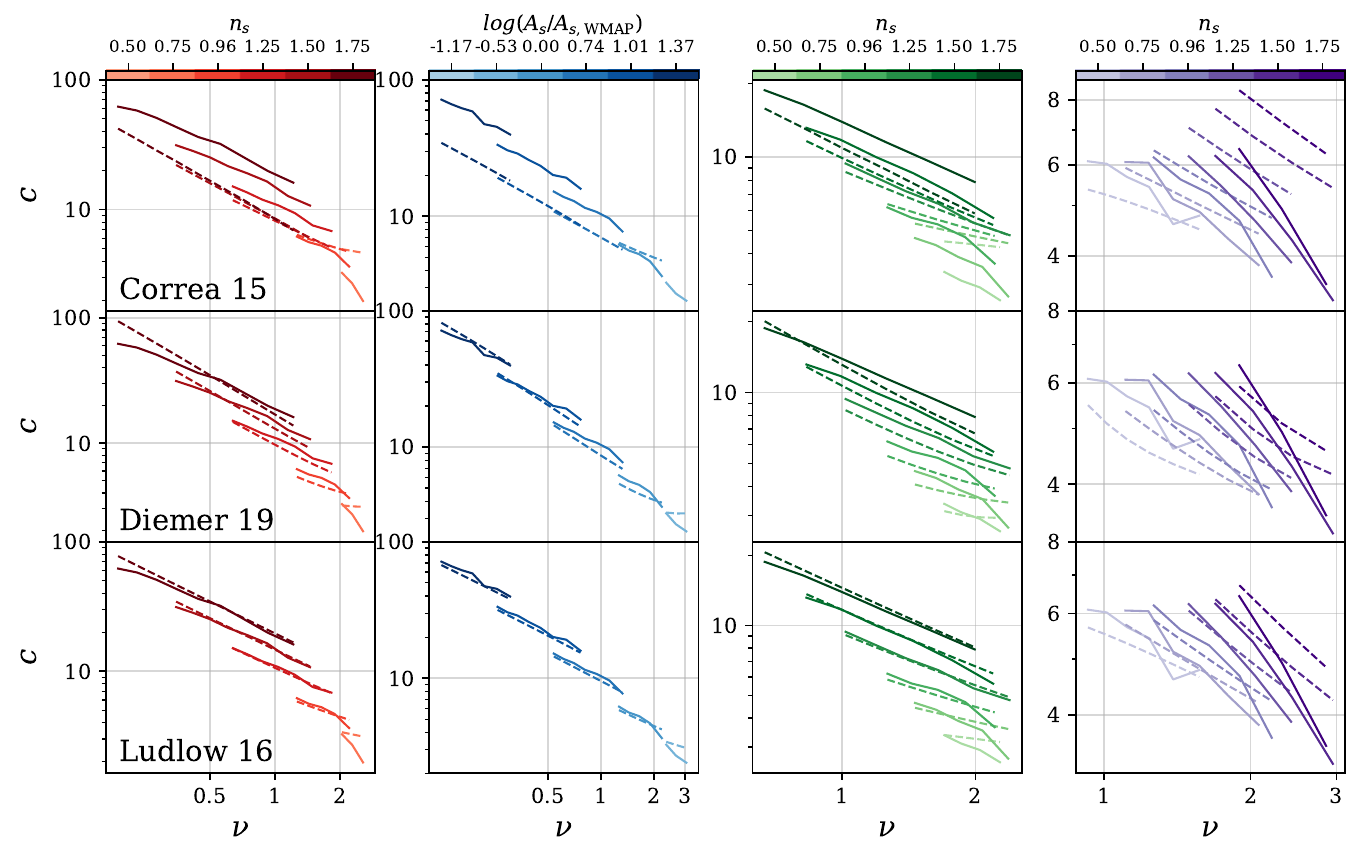}
    
    \caption{Comparisons between the predictions of \citet{correa_3}, \citet{Aaron_cm_relation_warm} and \citet{Diemer_2019} to this work for the $c\text{--} \nu $ relation. Predictions from the various works are shown as dashed lines while our results are solid lines. Each column represents the different simulations suites with the shaded colour the value of $n_s$ (or $A_s$). Each row is is the predictions from the different models, see bottom left of left most panel. Unlike Fig.~\ref{fig:overdensity_parameters} we have used variable dynamic ranges to more easily highlight the differences.}
    \label{fig:CM_prediction}
\end{figure*}

The model of Correa et al. predicts halo concentration given a mass and redshift for any general $\Lambda$CDM cosmology. There are two key parts to the model; the prediction of the MAH in a general cosmology using extended Press-Schechter theory and an empirical relation mapping a given MAH to a halo concentration found from simulations by \citet{Ludlow_CDM}. This empirical relation was found studying the Millennium simulations \citep{Millenium1,Millenium2,Millenium3}, which are at a fixed cosmology but variable resolution. We refer the reader to \cite{correa_1,correa_2,correa_3,Ludlow_CDM} for details.

In the top panels of Fig.~\ref{fig:CM_prediction} we present the predicted $c\text{--}\nu$ relation (dashed lines) for the model of Correa et al. compared to our results (solid lines). For the $n_s=0.96$ cosmology, the analytic model accurately predicts the $c \text{--} \nu$, as we would expect due to their work being tested and somewhat calibrated on a very similar cosmology. However, for cosmology distinctly different from our own the predictions consistently disagree from our results. In general it appears that for simulations with increased primordial amplitudes (darker red and blue lines) the concentration is under predicted while the opposite is true for the simulations with reduced primordial amplitudes. Studying the MAHs for the different cosmologies it appears that the prediction of Correa et al. don't agree with the results of this work (i.e. Fig.~\ref{fig:MAH}). In general it is found that for cosmologies with larger initial amplitudes the Correa et al. model predicts haloes to form later than we observe, vice versa for smaller initial amplitudes. The discrepancy in halo formation time is therefore in qualitative agreement with the differences in concentration observed in Fig.~\ref{fig:CM_prediction}. We re-emphasise that all cosmologies other than $n_s=0.96$ are well outside of observational constraints; as long as the predictions of these models are accurate for universes close to our own, which they appear to be, they can still be reliably applied/compared to observations.

\subsubsection{Comparison to the model of Diemer \& Joyce}
The work of \citet{Diemer_2015} studied scale free cosmologies, that being a universe with a power law linear power spectra, with differing slopes, $n$. They observed that at fixed $n$ the $c \text{--} \nu$ relation is universal, but does in general depend on the slope of the power spectrum. These results are therefore in general agreement to the key features observed in Fig.~\ref{fig:overdensity_parameters}; although $n$ and $n_s$ are not the same quantities they both represent a general `shape' change to the power spectra. The interpretation of this result was that the concentration of a halo depends on its peak-height as well as the effective slope of the linear power spectrum at an associated k-scale. From these observations a semi-analytic model was created and calibrated against simulations with a range of cosmologies to predict the $c\text{--}M$ in a general cosmology at any redshift or mass. The cosmologies used for calibration include WMAP 7-yr, Planck15 and scale free cosmologies. Here we compare to \citet{Diemer_2019} which is an updated version of the original \citet{Diemer_2015} model, we refer the readers to the papers for detailed differences between the models.

In the bottom panel of Fig.~\ref{fig:CM_prediction} we show the predictions of \citet{Diemer_2019} compared to our results for all simulations at $z=0$. We have shown here the $c\text{--}\nu$ relation instead of $c\text{--} M$ as the model fundamentally works with peak-height as oppose to mass. For the two suites dominated by amplitude changes (left and middle-left panels) the predictions of the model are in very good agreement with our results. There does appear to be a systematic amplitude offset from the predictions and our results. The method used to estimate concentration in our simulations is different to the one they used when calibrating their model. The observed amplitude offset is qualitatively consistent with the expectation from different methods of determining halo concentration as well as choosing whether to include a relaxation cut or not \citep[e.g.][]{Child_2018}. However, when focusing on the suite dominated by a shape change to the initial power spectra, we see that the predictions of Diemer et al.~do not reproduce the simulations particularly well. Specifically, their model does appear to match the amplitude but predicts a much flatter dependence on $\nu$ than we see in the simulations. 

\subsubsection{Comparison to the model of Ludlow et al.}
In the work of \citet{Aaron_cm_relation_warm} they demonstrated that the concentration of a halo, in both a cold and warm dark matter (WDM) universe, can be directly linked to its CMH. This was shown to work in a range of cosmologies, both with CDM and WDM cosmologies with differently assumed particles masses using the \texttt{COCO} \citep{coco_warm}, \texttt{Millenium} \citep{Millenium1,Millenium2,Millenium3} as well as some additional $\Lambda$CDM cosmologies. From this observation a semi-analytic model was created to predict halo concentration for a general cosmology.

In the middle panel of Fig.~\ref{fig:CM_prediction} we show the predictions of \citet{Aaron_cm_relation_warm} compared to our results. We see very similar results to model of Diemer \& Joyce; for cosmologies with higher concentrations, effectively higher initial amplitudes, the predictions match well our results. For cosmologies with lower concentrations, particularly the $k_{\rm{pivot}}=1 h$ Mpc$^{-1}$ suite, the predictions match the approximate magnitude but do not agree with the exact $\nu$ dependence. The model of Ludlow et al.~does not include a turn up in concentration, unlike the Diemer \& Joyce model, so the discrepancy at low concentration and large peak height appears to be independent of relaxation cuts. It is interesting that all models studied here essentially fail for this suite, it is not clear why this is the case and something we leave for future work.

\section{Summary}
\label{summary}

In this work we have examined how changes in the primordial power spectrum of density fluctuations affect the internal structure of haloes in a collisionless universe. We have done this by varying the amplitude of fluctuations, $A_s$, the primordial spectral index, $n_s$, and the normalisation $k$ scale (or pivot point), $k_{\rm{pivot}}$, within the context of a $\Lambda$CDM model with a fixed expansion history. By varying these parameters systematically (see Fig.~\ref{fig:initial_pow}), we are able to isolate the impacts of amplitude and shape variations in the initial power spectra on the properties of collapsed haloes. We find that when studying universes that deviate strongly from our own, some key results on the structure of dark matter haloes that have been assumed to be universal no longer hold true.  

The main results of our study are as follows:

(i) The mass structure of collapsed haloes retains a memory of the primordial power spectrum (see Fig.~\ref{fig:dens_profile}). It is found that the NFW form which works well for haloes in simulations with CMB-normalised fluctuations, breaks down when the amplitude of initial density fluctuations is increased (see Fig.~\ref{fig:NFW_fit}), these simulations correspond to haloes with small peak heights and early formation times. The NFW profile no longer offers a good description of halo density profiles in this regime due to their outer logarithmic slopes being steeper than $\gamma=-3$.  An Einasto form works relatively well for all of the simulations investigated here.

(ii) The pseudo-entropy (or pseudo-phase space density) profiles can be described relatively well by a simple power law in all cosmologies studied here (see Fig.~\ref{fig:entropy_profile}), however, the exponent is not a constant. A clear mass dependence is seen in many of the simulations (see Fig.~\ref{fig:mass_parameters}). For the case of CMB-normalised power spectra, there is only a very mild mass dependence of the range sampled by the simulations, with roughly a constant value of $\chi \approx 1.22$, in agreement with previous studies.

(iii) The general physical picture identified in many previous studies that the concentration of dark matter haloes is tied to halo formation time continues to hold in all the simulations examined here (see Fig.~\ref{MAH_CMH}). We find that the prediction for halo concentration of the models from \citet{Aaron_cm_relation_warm} and \citet{Diemer_2019} in general match well our results, particularly for cosmologies with much larger initial amplitudes to our own. However, in cosmologies with reduced primordial amplitudes, resulting in lower concentrations today, these models do not appear to accurately match our results predicting much shallower dependence on peak height, $\nu$, than we observe (see Fig.~\ref{fig:CM_prediction}).

(iv) It is found that the dominant effect on the density profiles of haloes, expressed through the concentration parameter, is  due to changes to the amplitude of the initial power spectra as oppose to the `shape' (see Fig.~\ref{fig:dens_profile} \& \ref{fig:mass_parameters}). The effect of that changing the amplitude of the primordial power spectra has on halo concentration is broadly encapsulated by peak height, $\nu$, but there are clear secondary effects (see Fig.~\ref{fig:overdensity_parameters}). However, it is found that changes to the shape of the initial power spectra (studied here through the primordial spectral index, $n_s$) are not accounted for by peak height alone. The secondary effects observed for only an amplitude change are exacerbated when the shape also changes, resulting in peak height correlating poorly with halo concentration for the suites dominated by a shape change to the linear power spectra (see right most panel of Fig.~\ref{fig:overdensity_parameters}). 

Summarising the above, our work has demonstrated that the internal properties (mass structure and dynamics) of collapsed haloes retain a clear memory of the initial conditions of the universe. We again point out that the simulations presented here are fully in the context of $\Lambda$CDM, in terms of expansion history and the nature of dark matter (cold and collisionless). Thus, our results indicate that the apparent universality of previously reported results for $\Lambda$CDM (e.g., NFW or similar forms for the mass structure, a power law of specific form for the phase-space density/pseudo-entropy profiles) is mostly a consequence of starting from a narrow range of normalisations for the initial power spectra.  Our work provides important new results that link the initial conditions to the present-day structure of collapsed dark matter haloes and can provide an important test-bench for physical models of structure formation.

\section*{Acknowledgements}
The authors thank the anonymous referee for a constructive report that improved the paper.  They also thank Aaron Ludlow, Joop Schaye, Camila Correa and particularly Simon White for providing invaluable feedback.
STB acknowledges an STFC doctoral studentship. This project has received funding from the European Research Council (ERC) under the European Union's Horizon 2020 research and innovation programme (grant agreement No 769130).
This work used the DiRAC@Durham facility managed by the Institute for Computational Cosmology on behalf of the STFC DiRAC HPC Facility. The equipment was funded by BEIS capital funding via STFC capital grants ST/P002293/1, ST/R002371/1 and ST/S002502/1, Durham University and STFC operations grant ST/R000832/1. DiRAC is part of the National e-Infrastructure.

\bibliographystyle{mnras}
\bibliography{References.bib}

\appendix 
\section{Density and velocity dispersion calculation methods} \label{appendix_dens_method}
Typically throughout the literature radial density profiles are calculated using logarithmically spaced bins. Essentially taking a histogram of the radial position of particles before normalising to convert to a mass density. Most often bins are be taken as a set number, $N_{\rm{bins}}$, spaced logarithmically from a minimum to a maximum fraction of $R_{\rm{200c}}$, $r_{\rm{min}}/R_{\rm{200c}} <r/R_{\rm{200c}} < r_{\rm{max}}/R_{\rm{200c}}$. Typical values are: $N_{\rm{bins}} \sim 20-50$, $log (r_{\rm{min}}/R_{\rm{200c}}) \sim -3 $ -- $ -1$ and $r_{\rm{max}}/R_{\rm{200c}}\sim 0.7-1$ (or much higher if the work specifically focused past $R_{\rm{200c}}$).

The standard approach described above is a particular form of a more general method to calculate the density (or similarly number/probability density) from a finite sample of data points. More generally a weight function can be used to estimate these quantities. There are two key free parameters associated with using a weight function, the `shape' of the function used and its width. In the standard method the `shape' is a top-hat function and the widths are chosen to be logarithmically spaced.

Associated with using a weight function are two types of error. The first being Poisson noise due to having a finite amount of data, this type of error is purely random. The second is systematic errors associated with the width of the kernel, essentially error due to trying to sample the density at a singular location by using data over a range of radii. These two errors work in opposite ways, Poisson errors decrease with a larger kernel width while the systematics respond in the opposite manner. This suggests that for a particular problem there is an optimal kernel width to use. 

Consider the following problem in one dimension. Given a set of particles (or simply data points) at positions $x_i$, where $i=1,2,3...N$, estimate the number density function, $n(x)$, that they have been sampled from. This is equivalent to trying to calculate the mass density as a function of radius for a halo, only with a different normalisation from 1D to 3D as well as a mass term. One way to estimate $n(x)$ is to use a weight function, as we will discuss here. To estimate $n(x)$ we use the following,
\begin{equation}
    n_{\rm{calc}}(x)=\sum W((x_i-x)/h).
\end{equation}
Here we have used the subscript $n_{\rm{calc}}(x)$ to distinguish this as our estimate of the true $n(x)$. $h$ is the `width' of the weight function and is a free parameter of the method. $W(x)$ is a general weight function which is assumed to have the following properties: (i) the function is symmetric about $x=0$ , (ii) the function is zero in the range $|x-x_i|/h>1$ and (iii) the weight function is normalised such that $\int_{-\infty}^{\infty}W(x)dx=1$. 

As discussed before, in general, $n(x) \neq n_{\rm{calc}}(x)$ due to two main types of error. We can therefore write, with a summation ansatz, $n(x) = n_{\rm{calc}} +n_{\rm{Poisson}}+n_{\rm{syst}}$. The first error term, $n_{\rm{Poisson}}$, is the Poisson error and completely random. We expect this error to scale as $n_{\rm{Poisson}} \propto \sqrt{N_{\rm{Kern}}}$, the number of points contained within the kernel. The second term is the systematic error, and is the residual left when $N_{\rm{Kern}} \to \infty$. The leading order dependence is as follows,
\begin{equation}
    n(x)+n_{\rm{sys}}=\int_{-\infty}^{\infty} n(x')W \bigg(\frac{x-x'}{h} \bigg) dx'=n(x)+\mathcal{O} (h^2 n''(x) ).
\end{equation}
The systematic error scales as $\mathcal{O} (n''(x)h^2$).

We cannot put a strict form on the optimal $h(x)$ as this would involve \textit{a priori} knowledge of $n''(x)$. We can, however, derive the scaling with the total number of data points $N_{\rm{tot}}$, i.e. how $h$ should behave in cases of different numerical resolution. If we require that for all resolutions we wish to be in a regime where neither error term dominates the other then we require that $n_{\rm{Poisson}}(N_{\rm{tot}}) \propto n_{\rm{sys}}(N_{\rm{tot}})$.

To proceed we need to parameterise the general form with which we will vary the kernel width. There are a few possible options, including using a fixed number of particles in the kernel.\footnote{Equivalent to how many smooth-particle hydrodynamics schemes estimate densities.} However, after trying a few different methods and comparing the results it appears the optimal option is choosing $h(x)$ such that it is equivalent to logarithmically spaced bins. The logarithmic spacing of bins can be calculated as the following,
\begin{equation}
    \Delta=(log_{10}(r_{\rm{max}})-log_{10}(r_{\rm{min}}))/N_{bins}
    \label{Delta}
\end{equation} 
and is related to the kernel width by 
\begin{equation}
    h(x)=(10^{\Delta}-1)/(10^{\Delta}+1)x=Ax.
    \label{h(x)}
\end{equation}
This highlights that the term `logarithmically spaced bins' is equivalent to a kernel width that scales linearly with radius. To first order it is found that $n_{\rm{Poisson}}=\sqrt{2n(x)h(x)}$ and $n_{\rm{sys}}= n''(x)h^2(x)$. The kernel width should therefore scale with $N_{\rm{tot}}$ in the following way,
\begin{equation}
\label{log_bins}
    A =h_0 N_{tot}^{-1/3} \quad \textrm{or} \quad \Delta = log \bigg (\frac{1+h_0 N^{-1/3}}{1-h_0 N^{-1/3}}\bigg) \approx \frac{h_0 N_{\rm{tot}}^{-1/3}}{\ln 10}.
\end{equation}
The parameter, $h_{0}$, depends in a complicated way on the particular weight function used and density function trying to be estimated. However, it is directly related to the ratio of systematic to random error, $n_{\rm{Poisson}}/n_{\rm{sys}}$, wished to be imposed. This derived behaviour with $N_{\rm{tot}}$ is consistent with the qualitative expectation; for a more highly sample function the kernel can be narrower to minimise systematics before counting error becomes dominant, and for more coarsely sampled profiles a wider kernel is required to have enough particles to reliably estimate $n(x)$.

An equivalent analysis can be applied to the velocity dispersion to find the equivalent scaling with $N_{\rm{tot}}$. Although the systematic errors have a different $N_{rm{tot}}$ dependence it is found that after imposing the condition that $n_{\rm{Poisson}}(N_{\rm{tot}}) \propto n_{\rm{sys}}(N_{\rm{tot}})$ that the same $N_{rm{tot}}$ dependence is required. We therefore also use $A =h_0 N_{tot}^{-1/3}$ for calculating velocity dispersions.

\subsection{Choosing appropriate parameters}
We can relate the above discussion directly to calculating the density and velocity dispersion profiles of real haloes in a cosmological simulation. Working in scale-free dimensionless coordinates, with radii scales by $R_{\rm{200c}}$, particle masses scales by $M_{\rm{200c}}$ and densities scaled by $\rho _{\rm{200}}$ then dark matter haloes are approximately self-similar. There is, however, a systematic dependence of the concentration parameter with mass that should be kept in mind. If, for the moment, we neglect this mass-concentration relation then what we would expect in such a self similar universe is the exact situation above were we want to estimate densities with a varying mass resolution (i.e. number of particles). For a density profile of finite mass $N_{\rm{tot}} \propto N_{\rm{200}}$, so we can directly use Eqn.~(\ref{log_bins}) with $N_{\rm{200}}$ in place of $N_{\rm{tot}}$. To be able to apply either method to a general N-body simulations we still need to find appropriate values of $h_0$. We will find the optimal choice of $h_0$ empirically using a numerical approach.

We use a Monte-Carlo approach to study how well various values of $h_0$ recover the density and velocity dispersion profiles. For this we study Einasto density profiles at variable resolutions. We assume that haloes are statistically perfect Einasto profiles and generate random positions such that they match the equivalent probability density function. The number density (per unit volume) of such a profile can be written in the following way,
\begin{equation}
    n(x)=B(\alpha,r_{-2})N_{\rm{200}}exp(-2/\alpha((x/x_{-2})^\alpha-1).
\end{equation}
$x=r/R_{\rm{200c}}$ (with $x_{-2}$ assumed to be similarly normalised) and $B$ is a normalisation factor, $B(\alpha,r_{-2})=1/(4\pi (x_{-2}/(2\alpha)^{1\alpha})^3 \Gamma (3\alpha) exp(2\alpha)  \Gamma(3\alpha,2/(\alpha x_{-2}^\alpha)))$. We then generate the probability density function by numerically solving for the cumulative distribution function.

In this analysis we try to recreate the process that would be applied to haloes as accurately as possible. We calculate the density and velocity dispersion at $30$ logarithmic spaced radii from $log(r)=-2 $--$ 0$. The figure of merit used to define the errors is as follows,
\begin{equation} \label{error}
    \psi^2=\sum_{\rm{N_{\rm{bins}}}}(log(\rho_{\rm{calc},i})-log(\rho))^2
    .
\end{equation}
We use this parameter to give an indication of the errors (both random and systematic) on the given profile. We also minimise this quantity to fit the calculated profiles. All profiles are only fit over the range where the sampled radii exceed the convergence radius \citep{Convergence}. 

We also study how these quantities vary with different concentration parameters. We assume $\alpha=0.16$ for all of the subsequent analysis. For each quantity we look at how errors vary with $N_{\rm{200c}}$, concentration (assuming $\alpha=0.16$) as well as the method employed. We also include a fiducial model where the standard method is used with $31$ logarithmically spaced bins over the same range, with the key point being that it is a method whose width does not vary as a function of $N_{\rm{200c}}$. 

For each $N_{\rm{200c}}$ we average over $100$ different random seeds to obtain typical values. As well as systematically varying $x_{-2}$ (from $1/2$ to $1/50$).

\begin{figure}
    \centering
    \includegraphics[width=\columnwidth]{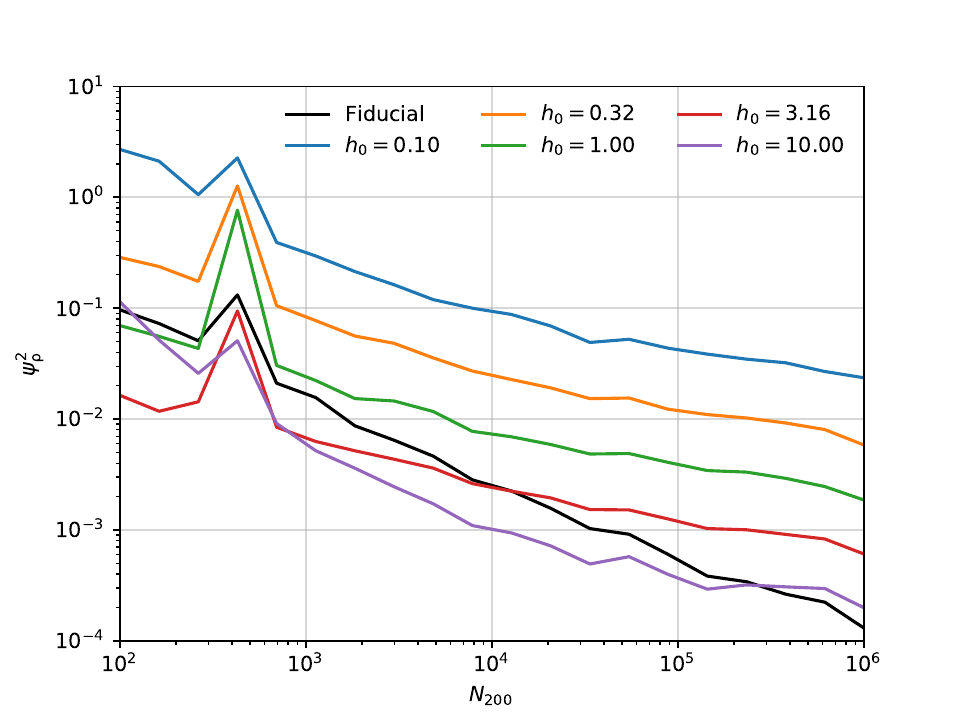}
    \caption{Figure of merit for the calculated density profiles, see Eqn.~\ref{error}, as a function of resolution, $N_{\rm{200c}}$. Each value is averaged over $100$ different haloes. Each line represents a different choice of parameter, $h_0$, see legend. Plotted in black on each plot is the standard method.}
    \label{error_density}
\end{figure}

In Fig.~\ref{error_density} we present how $\psi^2$ (averaged over 100 different seeds) varies as a function of $N_{\rm{200c}}$ for a halo with $r_{-2}=1/10$ and $\alpha=0.18$. In general the wider the kernel (larger values of $h_0$) results in a smaller $\psi^2$. The only exception to this is in the region $N_{\rm{200c}}<10^3$ which, for $h_0=10$, corresponds to the where $h>r$ so will end up with non-physical results. Looking at the fiducial model (black line) it appears that it is a middle ground to some of the other parameters, with errors at low $N_{\rm{200c}}$ larger than other models but performing better at high values. The fiducial model appears to be the optimal choice in regard to minimising $\psi^2$ at $N_{\rm{200c}}=10^6$, however for most simulations haloes of this size will only be a few and the errors for $h_0=10$ and $3.16$ are still small.

\begin{figure}
    \centering
    \includegraphics[width=1\columnwidth]{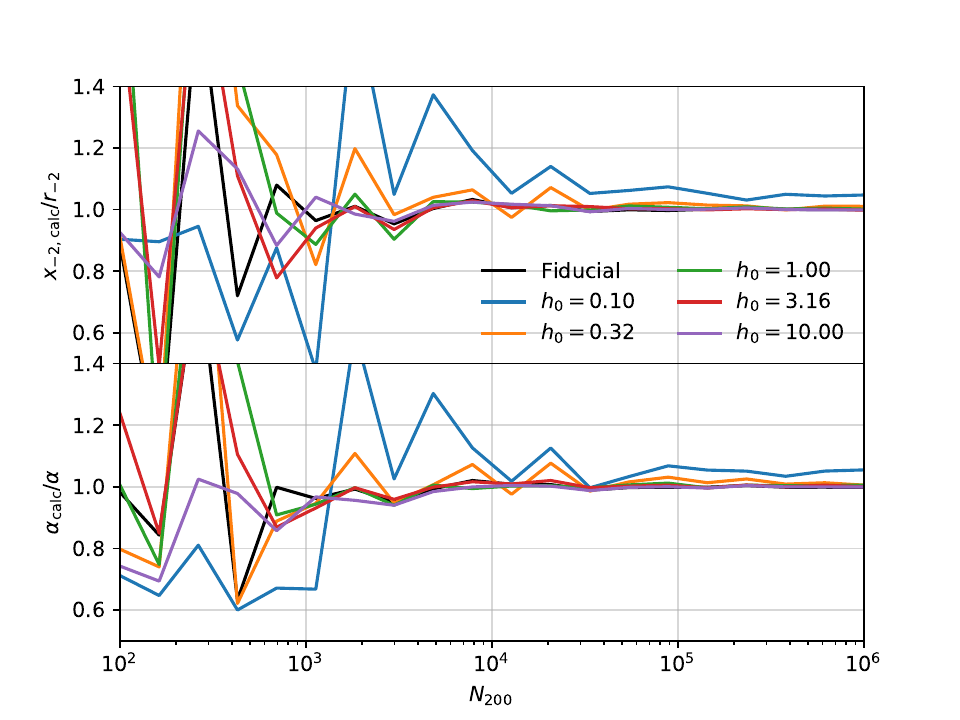}
    \caption{Fitted Einasto parameters, $x_{-2}$ (top panel) and $\alpha$ (bottom panel), as a function of resolution for the various sampled parameters, $h_0$, see legend. The fitting is done to minimise $\psi^2$ and is fit over the range $r_{\rm{conv}}<r<R_{\rm{200c}}$, where $r_{\rm{conv}}$ is the convergence radius, and the values are averaged over $100$ different haloes. In each panel the standard method is plotted in black for reference.}
    \label{paramater_density}
\end{figure}

We next focus on the averaged fitted parameters, $x_{-2}$ and $\alpha$, to make sure they can be reliably recovered. This part of the analysis should most definitely be considered as it was found that minimising $\psi^2$ does not necessarily minimise the error on the fit and can, if mismanaged, lead to significant systematic errors. We present the average of these fitted parameters in Fig.~\ref{paramater_density}. We can see that for the well resolved haloes, $N_{\rm{200c}}>10^3$, All choices of parameters other than $h_0=0.1$ results in minimal systematic on the fitted parameters with errors $\sim 1\%$.

Based on this analysis we it appears that the best choice is $h_0=10$. However this analysis has been done on \textit{mathematically perfect} haloes. For real haloes with substructure we want a kernel narrower than the optimal one found from this study. We therefore choose $h_0=500^{1/3}$. We also apply a resolution cut of $N_{\rm{200c}}=2000$ so we are comfortable in the regime where $\psi^2<10^-2$ and the large scatter found in fitted parameters in the low resolution limit ($N_{\rm{200c}}<10^3$).

Although not presented here, the analysis was also applied to haloes with concentrations ranging from $c=2$ to $50$. The key results are the same and doesn't change the conclusion of the optimal value of $h_0$. The only notable difference observed for differently concentrated haloes is that haloes with lower concentrations lead to larger errors both on the density profiles and the fits. This trend is fairly mild, but does rule out issues with calculating the densities for the extreme concentrations observed in this work. There were no observed differences in the reliability of finding the densities for variation in $\alpha$.

All of the above analysis was applied to the velocity dispersions as well. To generate a velocity dispersion we numerically solve the Jeans equation, assuming isotropic velocities, to find the velocity dispersion as a function of radius for a given Einasto profile normalising it so that the maximum velocity dispersion is unity. The velocities are then assumed to be a Gaussian about zero with a standard deviation equal to the velocity dispersion. Although it didn't have to be the case it is found that $h_0=500^{1/3}$ is also optimal.

\section{Resolution test and box size tests} \label{Resolution_test}
In this work we have assumed that the density, entropy and velocity dispersion profiles are converged in the regime where $r>r_{\rm{conv}}$. Where $r_{\rm{conv}}$ is the found using the relation in \cite{Convergence}. In the work of \cite{Convergence} the condition for convergence was derived from a suite of simulations with cosmologies close to our own. As such the haloes studied were drastically different to some in this work, with orders of magnitude different concentrations. It is therefore possible that this condition for convergence does not hold in these more extreme cosmologies.

To check that this is not a factor in our work we have two different resolution simulations WMAP9 pivot with variable $n_s$ suites with a box size of $100h^{-1}$Mpc. In this way we can make sure that both amplitude and shape changes to the initial power spectra are not a convergence issue. The two comparison suites of simulations are almost identical (same box size, same cosmologies, same initial redshift, etc) but use $512^3$ and $256^3$ particles, corresponding to particle masses of $4.62 \times 10^{9}h^{-1}$ M$_{\odot}$ and $5.78 \times 10^{8}h^{-1}$ M$_{\odot}$ respectively.\footnote{The softening length has also been changes appropriately from $\epsilon=4 h^{-1}$ Mpc to $\epsilon=2 h^{-1}$ Mpc.} We have modified the scaling relations for the kernel  width, Eqn.~(\ref{kernel_width_scaling}), to make the different resolution simulations equivalent and remove this as a factor in the comparison.

\begin{figure}
    \centering
    \includegraphics[width=\columnwidth]{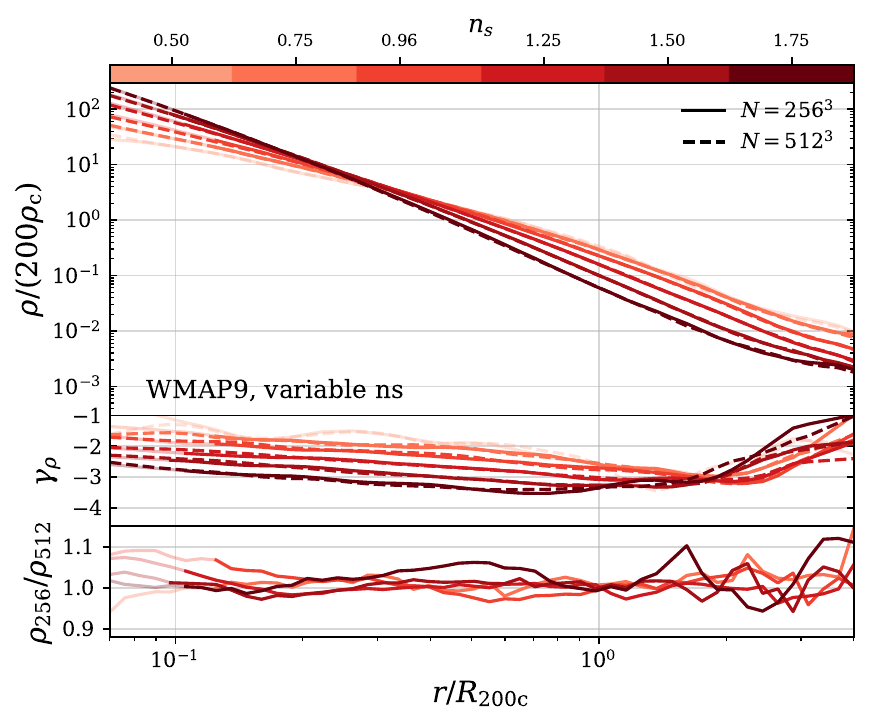}
    \caption{Stacked density profiles for haloes in the mass range $M_{\rm{200,c}}=10^{13}$--$10^{13.5}h^{-1}$ M$_{\odot}$ for two different mass resolution simulations. Solid lines represent the simulation using $256^3$ particles while dashed $512^3$. The different shades represent the different values of $n_s$ used to generate the initial conditions. In the top panel is shown the density profiles, the middle panel shows the logarithmic slope while the bottom panel the density ratios for the two different resolutions. Transparent parts of the lines represent regions not meeting the condition for convergence, while opaque lines represent the opposite.}
    \label{density_res_test}
\end{figure}

In Fig.~\ref{density_res_test} we present the stacked density profiles for haloes between $10^{13} h^{-1}$ M$_{\odot}$ and $10^{13.5} h^{-1}$ M$_{\odot}$, hence directly comparable to Fig.~\ref{fig:dens_profile}, of the two different resolution suites of simulations. As with the rest of the paper different shades of red represent different values of $n_s$ while here the two different line styles show the different resolutions, see legend. In the top panel is plotted the density, the middle panel the logarithmic slope and the bottom panel show the ratio of densities for the two different resolutions. 

It can be seen in the ratio plots that all of the simulations studies here are indeed converged with $r>r_{\rm{conv}}$. With maximum errors of $\sim 10 \%$ and more typical errors of only a few percent. Although not shown here we see equivalent convergence trends for the entropy and velocity dispersion profiles. In the regime $r>r_{\rm{conv}}$ they are well converged, again $<10\%$. 

\begin{figure}
    \centering
    \includegraphics[width=\columnwidth]{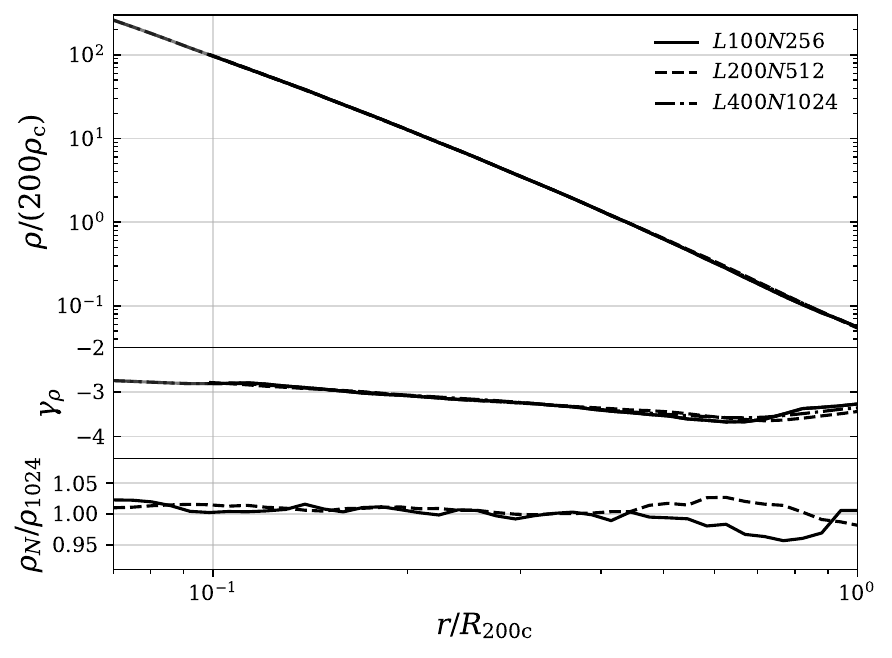}
    \caption{Stacked density profiles for haloes in the mass range $M_{\rm{200,c}}=10^{13}$--$10^{13.5}h^{-1}$ M$_{\odot}$ for the most extreme cosmology in the suite with variable $A_s$, corresponding to $\sigma_8=8.103$ (see Table~\ref{simulation_table}), simulated at three different box sizes with fixed mass resolution. Solid lines represent the simulation using $256^3$ particles and a box size of $100$ h$^{-1}$Mpc, dashed lines represent $512^3$ particles with a box size of $200$ h$^{-1}$Mpc and dot-dashed lines represent $1024^3$ particles with a box size of $400$ h$^{-1}$Mpc. The top panel show the densities, the middle the logarithmic slope and the bottom panel the ratio with respect to the largest box.}
    \label{fig:density_box_test}
\end{figure}

Additionally we have checked that the finite box size is not an issue for the statistics studied in this work. This was done by simulating the most extreme cosmology in the suite with variable $A_s$, corresponding to $\sigma_8=8.103$ (see Table~\ref{simulation_table}) at three different box sizes with consistent particle masses. The results of this test are shown in Fig.~\ref{fig:density_box_test}, where it can be seen that the density profiles are converged to within a few percent for all box sizes studied here. Although not shown here we find similarly well converged results for the entropy and velocity dispersion/anisotropy profiles.

\label{lastpage}
\end{document}